\documentclass{jfm}
\usepackage{amsmath}
\usepackage{amsfonts}
\usepackage{amssymb}
\usepackage{graphicx}
\usepackage[round]{natbib}
\usepackage{float}
\usepackage{epstopdf}

\usepackage{lineno}
\usepackage{multirow}
\usepackage{color}

\newcommand*\patchAmsMathEnvironmentForLineno[1]{%
  \expandafter\let\csname old#1\expandafter\endcsname\csname #1\endcsname
  \expandafter\let\csname oldend#1\expandafter\endcsname\csname end#1\endcsname
  \renewenvironment{#1}%
     {\linenomath\csname old#1\endcsname}%
     {\csname oldend#1\endcsname\endlinenomath}}% 
\newcommand*\patchBothAmsMathEnvironmentsForLineno[1]{%
  \patchAmsMathEnvironmentForLineno{#1}%
  \patchAmsMathEnvironmentForLineno{#1*}}%
\AtBeginDocument{%
\patchBothAmsMathEnvironmentsForLineno{equation}%
\patchBothAmsMathEnvironmentsForLineno{align}%
\patchBothAmsMathEnvironmentsForLineno{flalign}%
\patchBothAmsMathEnvironmentsForLineno{alignat}%
\patchBothAmsMathEnvironmentsForLineno{gather}%
\patchBothAmsMathEnvironmentsForLineno{multline}%
}

\title[Interactions between mountain waves and shear flow]{Interaction between mountain waves and shear flow in an inertial layer}
\author[J.-H. Xie \& J. Vanneste]{Jin-Han Xie$^1$ and Jacques Vanneste$^2$}
\affiliation{$^1$ Department of Physics,
University of California, Berkeley, CA 94720, USA \smallskip \\ $^2$ School of Mathematics and Maxwell Institute for Mathematical Sciences, \\
University of Edinburgh, Edinburgh EH9 3FD, UK}

\newcommand{\av}[1]{\left<{#1}\right> }

\newcommand{\mb}[1]{\mathbf{#1}}
\newcommand{\mr}[1]{\mathrm{#1}}
\newcommand{\br}[1]{\left( #1 \right)}
\newcommand{\ex}{\mathrm{e}}
\newcommand{\ii}{\mathrm{i}}
\newcommand{\Ro}{\textrm{Ro}}
\newcommand{\Ri}{\textrm{Ri}}

\newcommand{\dd}{\mathrm{d}}
\newcommand{\bx}{\boldsymbol{x}}
\newcommand{\bk}{\boldsymbol{k}}
\newcommand{\bu}{\boldsymbol{u}}

\newcommand{\bU}{\boldsymbol{U}}
\newcommand{\bF}{\boldsymbol{\mathcal{F}}}

\newcommand{\tk}{k_*}
\newcommand{\tl}{l_*}
\newcommand{\btk}{\boldsymbol{k}_*}

\newcommand{\tz}{z_*}

\newcommand{\ttimes}{\!\times\!}
\renewcommand\Re{\mathrm{Re}\,}

\begin{document}
\maketitle

\begin{abstract}
	
	Mountain-generated inertia-gravity waves (IGWs) affect the dynamics of both the atmosphere and the ocean through the
	mean force they exert as they interact with the flow. A key to this interaction is the presence of critical-level singularities or, when planetary rotation is taken into account, inertial-level singularities, where the Doppler-shifted wave frequency matches the local Coriolis frequency. We examine the role of the latter singularities by studying the steady wavepacket generated by a multiscale mountain in a rotating linear shear flow at low Rossby number. Using a combination of WKB and saddle-point approximations, we provide an explicit description of the form of the wavepacket, of the  mean forcing it induces, and of the mean-flow response. 
	
	We identify two distinguished regimes of wave propagation: Regime I applies far enough from a dominant inertial level for the standard ray-tracing approximation to be valid; Regime II applies to a thin region where the wavepacket structure is controlled by the inertial-level singularities. 
	The wave--mean-flow interaction is governed by the change in  Eliassen--Palm  (or pseudomomentum) flux. This change is localised in a thin inertial layer where the wavepacket takes a limiting form of that found in Regime II. 
	We solve a quasi-geostrophic potential-vorticity equation forced by  the divergence of the Eliassen--Palm flux to compute the wave-induced mean flow. Our results, obtained in an inviscid limit, show that the wavepacket reaches a large-but-finite distance downstream of the mountain (specifically, a distance of order {$(k_*\Delta)^{1/2} \Delta$}, where $k_*^{-1}$ and $\Delta$ measure the wave and envelope scales of the mountain) and extends horizontally over a similar scale.

\end{abstract}

\section{Introduction}

The importance of mountain-generated inertia-gravity waves for the atmospheric circulation has long been recognised \citep[see][for a review]{Frit2003}, and their parameterisation is now an essential element of weather-forecasting and climate models \citep[e.g.,][]{Alex2010}. Their oceanic counterparts, while often neglected, are now increasingly thought to  play a significant role for the oceanic circulation \cite[e.g.][]{Scot2011,Niku2011,Niku2013}. These waves impact both the atmospheric and oceanic circulations through the drag they extert where they dissipate, often through their interaction with the large-scale flow at critical levels where the mean flow velocity vanishes or, accounting for the background rotation, at inertial levels where their Doppler-shifted frequency matches the local Coriolis frequency. %\commentJHX{Maybe delete the bracket since what in it is the information we stress.}

Our understanding of this form of interaction with the mean flow rests on a number of now classical papers \citep[including][]{Elia1961, Bret1966,Bret1969,Bret1969b,Jones1967,Book1967} that tackled both the propagation of the waves in a shear flow and the drag they exert on the flow. These identified the Eliassen--Palm (EP) flux (or pseudomomentum flux) as the key quantity controlling the drag, showed that its conservation in the absence of dissipation leads to non-interaction results \citep{Char1961,Andr1976,Andr1978}, and elucidated how
critical-level and inertial-level singularities disrupt this conservation and result in drag.  
These results have subsequently been applied to a variety of mountain shapes and flows.

The present paper focusses on the case of a topographic profile with two well-separated horizontal scales, with small-scale oscillations modulated over a large envelope scale. Topographies of this form are assumed in atmospheric-model parameterizations \citep{Mart2007} and are natural for the ocean, e.g., in ridge regions. Our aim is to provide a detailed description of the wavepacket  generated by a relatively weak flow whose Rossby number based on the envelope scale is small. The Rossby number based on the oscillation scale is however large enough for the waves to be vertically propagating (rather than evanescent) from the ground up. The flow considered is back sheared, decreasing linearly from a positive value at the ground to a zero-velocity critical level higher up.
Such a monotonic decrease is not particularly realistic, since real flows are more typically non-monotonic and critical levels, e.g. in the middle atmosphere, are the result of shear reversals. However, the important wave dynamics that we intend to capture is localized around the inertial levels and unaffected by the details of the propagation below, which are well described by standard ray tracing.

Background rotation is crucial in two respects for the problem considered: first, it contributes to the dispersion relation; second, it determines the nature of the singularities in the vertical structure of the wave solution. Specifically, rotation resolves the degeneracy of the critical level singularity, which is independent of wavenumber, into a pair of wavenumber-dependent inertial levels. As a result, the singularities associated with the broad wavenumber spectrum of a wavepacket are smeared out over a range of altitude -- the inertial layer -- and the wavepacket solution itself is smooth in the limit of vanishing dissipation \citep{Shut2001}. (An analogous effect arises when the orientation of the flow changes with altitude, see \citet{Shut1995,Shut2003}.)

We tackle the three essential aspects of the problem by computing (i) the shape of the wavepacket, (ii) the associated EP flux, and (iii) the mean-flow change that results from the divergence of this EP flux. 
We take advantage of the assumption of small Rossby number and of a related assumption of large Richardson number to carry out the entire computation asymptotically, relying on the WKB form of the vertical structure of plane waves in the horizontal obtained by \citet{Lott2010,Lott2012}. The analysis identifies two distinct altitude ranges corresponding to two distinct asymptotic regimes. In the first, valid away from the inertial levels, standard ray tracing applies and the (horizontally integrated) EP flux is independent of altitude; in the second, valid in a thin region surrounding the inertial layer, the solution is more complicated and captures the finiteness of the wavepacket deflection as it approaches a central inertial level. It turns out that inertial-level absorption affects only a still thinner region, which defines an inertial layer. The mean drag is vertically localised in this layer.

The asymptotic approach provides answers to basic questions -- such as the horizontal distance between mountain and  region of wave drag, and the extent of this region -- as scaling laws in terms of key parameters characterising the stratification, shear and mountain shape. These scaling relations, which likely apply to more general setups than the one we consider, may prove useful for the representation of moutain-wave drag in numerical models.
{We also emphasise that our approach provides a fully consistent treatment of wave--mean flow interaction in a non-symmetric setup, with small-scale averaging replacing the more familiar zonal averaging. This is in contrast with the earlier, heuristic treatment of \citet{Mart2007}.}

The structure of this paper is as follows. We formulate the problem in \S\ref{Formulation} and 
approximate the form of the wavepacket in different regimes using a steepest-descent method in  \S\ref{wave_solu}. 
In  \S\ref{Sec_wm} we use this approximation to calculate the EP flux and  solve a mean quasi-geostrophic potential-vorticity equation to obtain the mean-flow response.
We summarise and discuss our results in \S\ref{conclusion}.

\section{Formulation} \label{Formulation}

\begin{figure}
	\centering
	\includegraphics[width=\textwidth]{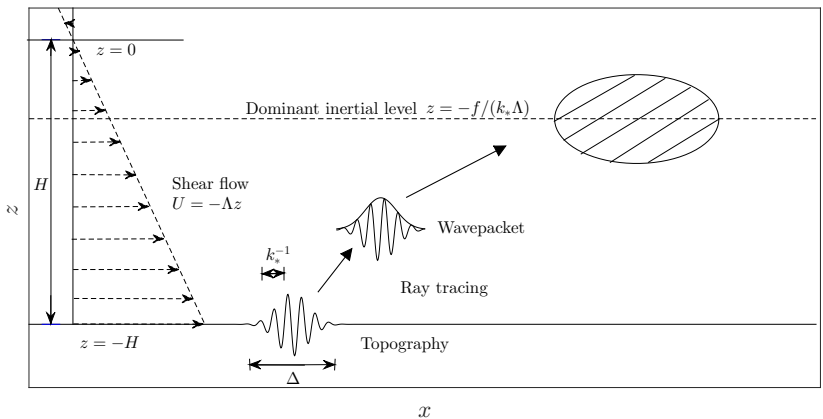}
	\caption{
		Schematic representation of the setup of the problem in the $(x,z)$-plane.
		A wavepacket generated by a two-scale mountain at the ground $z=-H$ propagates vertically in the shear flow $\bU = (-\Lambda z,0)$ and drives a mean-flow in the (hatched) region localised around the dominant inertial level $z=-f/(k_* \Lambda)$.
	}\label{Fig_setup}
\end{figure}

We consider the interaction between a steady topographic wavepacket and a background shear flow in an idealized setup shown in Fig.\ \ref{Fig_setup}. The background flow is chosen as a unidirectional, uniform backward shear flow $\bU = (-\Lambda z,0,0)$ with $\Lambda=\mathrm{const.}>0$. The distance between the level of zero background velocity (critical level) and the bottom boundary is $H$. 
It proves convenient to use a slightly unusual vertical coordinate such that $z = 0$ and $z = -H$ correspond to the critical level and ground, respectively. 
The topographic wavepacket is generated by an idealized multiscale mountain with height
\begin{eqnarray}
h_\mathrm{t}(x,y) &=& h \, \Re\left( \ex^{-(x^2+y^2)/(2\Delta^2)}\ex^{\ii \btk\cdot \bx} \right) \nonumber\\
&=& \frac{\Delta^2 h}{2\pi} \Re \left( \int_{-\infty}^{\infty} \int_{-\infty}^{\infty} \ex^{-|\bk-\btk|^2\Delta^2/2}\ex^{\ii \bk \cdot \bx } \dd k \dd l \right), \label{mountain}
\end{eqnarray}
where $\bx=(x,y)$, $\bk=(k,l)$, $\bk_*=(k_*,l_*)$ is the dominant wavevector, $h$ is the maximum height of the mountain, and $\Re$ denotes the real part. Here $k_*^{-1}$ and $\Delta$ control the oscillation scale and envelope scale of the moutain so that the parameter $k_*\Delta \gg 1$ characterizes the separation between these scales.

The fluid satisfies the $f$-plane hydrostatic Boussinesq equations
\begin{subequations}\label{PriEqu}
	\begin{align}
	\partial_t \bu + \bu\!\cdot\!\nabla\bu + w\partial_z \bu + f\boldsymbol{e}_{z} \!\times\! \bu &=- \nabla \phi, \label{Primomh}\\ 
	\partial_z \phi &= b, \\
	\partial_t b + \bu\!\cdot\!\nabla b + w\partial_z b + N^2 w &= 0, \label{Pript} \\
	\nabla\!\cdot\!\bu + \partial_z w &= 0,  \label{Priincomp}
	\end{align}
\end{subequations}
where $\bu=(u,v)$ is the horizontal velocity,
$w$ the vertical velocity,
$\phi$ a scaled pressure, $b$ the buoyancy,
$f$ the local Coriolis frequency,
$\boldsymbol{e}_{z}$ the unit vertical vector pointing upwards, $N$ the Brunt-V\"{a}is\"{a}l\"{a} frequency, taken to be a constant, and $\nabla = (\partial_x, \partial_y)$ is the horizontal gradient.

We apply a no-normal flow boundary condition at the lower boundary:
\begin{eqnarray}
w = \bu_{\mr{b}}\!\cdot\!\nabla h_\mathrm{t} \quad \mathrm{at} \quad z=-H+h_\mathrm{t}, \label{bc}
\end{eqnarray}
where the subscript ``$\mr{b}$'' denotes the value on the boundary.

\section{Wave solution} \label{wave_solu}

\subsection{Preliminaries} \label{sec_wave_pre}

We examine small-amplitude waves governed  by the linearization of the primitive equations (\ref{PriEqu})--(\ref{bc}) about the background flow. This is in geostrophic balance and given by
\begin{equation}
U_0 = -\partial_y \Psi_0, \quad V_0 = 0, \quad W_0=0, \quad B_0=f\partial_z \Psi_0, \quad \Phi_0 = f\Psi_0, \quad \mathrm{where} \quad \Psi_0 = \Lambda y z. \label{bgflow}
\end{equation}
We assume a small Rossby number  based on the envelope scale  $\Delta$ of the topography:
\begin{equation}
\Ro=\frac{U_{\mr{b}}}{f\Delta} \ll 1,
\label{smallRossby}
\end{equation} 
where $U_\mathrm{b} = \Lambda H$. {We also assume that $NH/(f \Delta) = O(1)$, corresponding to an order-one Burger number based on the horizontal and vertical scales $\Delta$ and $H$.
Together with (\ref{smallRossby}), this implies a large Richardson number, specifically
\begin{equation}
J =  \frac{N}{\Lambda} = O (\Ro^{-1}) \gg 1,
\label{Jscale}
\end{equation}
where we have introduced the parameter $J = \Ri^{1/2}$ as a convenient substitute for the Richardson number $\Ri$.}
%\commentJV{Swapped small-Rossby condition and background flow: this flow is an exact solution and would be OK for large Rossby.}
%of the mean flow\footnote{
%As we will see in the wave-mean flow interaction part that the mean flow generation can be described by the quasi-geostrophy theory.
%so in fact $\Delta$ is regarded as the scale of the mean flow generation, but this does not harm the mean shear flow which is an exact solution of primitive equations.
%},  

We are interested in mountain waves, in the distinguished regime where the Doppler-shifted frequency $U_\mathrm{b} k_*$ is of the same order as the Coriolis frequency. This corresponds to the scaling
{
\begin{equation}
r = \frac{U_\mathrm{b} k_*}{f} = O(1),
\label{ro}
\end{equation}
with $r$ a Rossby number based on the mountain wavelength rather than the envelope scale that appears in $\Ro$. 
Since $r = k_* \Delta \Ro$,  this implies that 
\begin{equation}
k_* \Delta = O(\Ro^{-1}) \gg 1.
\label{kDscale}
\end{equation}
To ensure that the waves are propagating rather than evanescent at the bottom boundary we further require that
\begin{equation}
r  > 1.
\label{prop}
\end{equation}
We emphasise that the scaling specified by (\ref{smallRossby}), (\ref{Jscale}) and (\ref{kDscale}) corresponds to a distinguished limit, that is, it leads to results valid for a broad range of relative values of $\Ro$, $J$ and $k_* \Delta$. This becomes apparent in the asymptotic derivation of \S\ref{sec:saddles} and is discussed further in the Conclusion, \S\ref{conclusion}.}

Linearizing (\ref{PriEqu}) around the background flow (\ref{bgflow}) leads to the equations
\begin{subequations}\label{WavePri}
	\begin{align}
	\partial_t \bu_1 - \Lambda z\partial_x\bu_1 + \Lambda w_1 \boldsymbol{e}_{x}  + f\boldsymbol{e}_{z} \!\times\! \bu_1 &=- \nabla \phi_1,  \label{WavePri_mom} \\ 
	\partial_z \phi_1 &= b_1, \\
	\partial_t b_1 - \Lambda z\partial_x b_1 + f\Lambda v_1 + N^2 w_1 &= 0,  \\
	\nabla\!\cdot\!\bu_1 + \partial_z w_1 &= 0.  \label{wave_incom}
	\end{align}
\end{subequations}
with $\boldsymbol{e}_{x}$ the unit vector in the $x$-direction,
which govern the leading-order wave fields, denoted here by the subscript ``1''.
The amplitude of the waves is determined by the linearisation of the boundary condition (\ref{bc}) {around the background flow (\ref{bgflow})},
\begin{equation}
w_1 = U_\mathrm{b} \partial_x h_\mathrm{t}.  \label{w_bc}
\end{equation}
This indicates that $w_1 = O( U_\mathrm{b}k_*h)$. The polarization relation for standard internal waves (see also (\ref{hatu}) below) can then be used to estimate $ u_1 = O( N w_1/(k_*U_\mathrm{b})) = O(Nh)$. 
The linearization based on the small-amplitude condition $u_1\ll U_\mathrm{b}$ thus requires that the inverse Froude number {be small}:
\begin{equation}
\frac{N h}{U_\mathrm{b}} = J\frac{h}{H} \ll 1. \label{smallpara}
\end{equation}
%where $J =N/\Lambda=\Ri^{1/2}$, with $\Ri=N^2/\Lambda^2$ the Richardson number.
Based on this small parameter, we introduce the convention of using a subscript `$n$' ($n=1,2,3\dots$) to denote the $n$th-order flow variables such that
\begin{equation}
u_n = O\br{ \br{J\frac{h}{H}}^n U_\mathrm{b} }.
\end{equation}

After applying a Fourier transform, (\ref{WavePri}) can be reduced to the single equation
\begin{equation}
\frac{1-\zeta^2}{\zeta^2}\hat{w}_{\zeta\zeta} - \br{\frac{2}{\zeta^3}-\frac{2\ii \nu}{\zeta^2}}\hat{w}_{\zeta} - \br{\frac{(1+\nu^2)J^2}{\zeta^2}+\frac{2\ii \nu}{\zeta^3}}\hat{w} = 0, \label{WaveEq}
\end{equation}
where $\nu = l/k$. (This is equation (25) in \citet{Jones1967} and equation (4) in \citet{Yama1984}.) The independent variable is the scaled vertical coordinate 
\begin{equation}
\zeta = - k\Lambda z/f.
\end{equation}
We emphasise that for $k>0$, as will be assumed when interpreting the results, $\zeta$ has a sign opposite to that of $z$ and is positive below the inertial level and increasing downwards.
The dependent variable $\hat{w}$ is the horizontal Fourier transform of $w_1$, defined by
\begin{equation}
w_1{(x,y,z)}= \frac{1}{2\pi}\int_{-\infty}^{\infty}\int_{-\infty}^{\infty} \hat{w}{(k,l,z)} \ex^{\ii \bk\cdot \bx } \, \dd k \dd l.
\label{hatwdef}
\end{equation} 
The other dependent variables are related to $\hat{w}$ through the polarization relations
\begin{subequations}\label{PolRel}
	\begin{align}
	\hat{u} &= \ii\frac{-\Lambda}{f}\br{ \frac{\zeta - \ii \nu}{\zeta(1+\nu^2)} \hat{w}_{\zeta} + \frac{\nu^2}{\zeta(1+\nu^2)} \hat{w} }, \label{hatu}\\
	\hat{v} &= \frac{\Lambda}{f}\br{ \frac{1 - \ii \nu\zeta}{\zeta(1+\nu^2)} \hat{w}_{\zeta} + \frac{\ii\nu}{\zeta(1+\nu^2)} \hat{w} },\\
	\hat{b} &= \ii\frac{\Lambda^2}{f}\br{ \frac{1 - \ii \nu\zeta}{\zeta^2(1+\nu^2)} \hat{w}_{\zeta} + \br{ \frac{\ii \nu}{\zeta^2(1+\nu^2)}  + \frac{J^2}{\zeta}}  \hat{w} }. \label{hatb}
	\end{align}
\end{subequations}

One of the key characteristics of IGWs in shear flow is the presence of singularities: two inertial levels, where the Doppler-shifted frequency matches the Coriolis frequency, and a critical level, where the Doppler-shifted frequency and hence the background velocity vanish \citep{Jones1967}.
These singularities are readily identified from (\ref{WaveEq}): the two inertial levels and one critical level correspond to $\zeta = \pm 1$ and $\zeta = 0$, respectively. 
The critical level is an apparent singularity that can be removed by a variable transformation. The inertial levels, by contrast, have a marked physical impact since the wave solution switches abruptly from an oscillatory to an evanescent behaviour and  back across them  \citep{Yama1984,Lott2015}. As we discuss in \S \ref{Sec_wm}, this abrupt change underpins the forcing of a  mean flow by the wavepacket.

Rotation plays a crucial role. The position $z=\pm f/(k \Lambda)$ of the inertial levels depends on the wavenumber $k$; 
%\commentJHX{I want to change the previous sentence to: Rotation, which leads to the $k$ dependence of the inertial level position $z=\pm f/(k \Lambda)$, plays a crucial role. As...}
as a result, the singularities  associated with each wavenumber making up the wavepacket are smeared out over a range of altitude, and the wavepacket solution is smooth even in the absence of  dissipation (or more precisely in the limit of vanishing dissipation since dissipation is important to determine  physically relevant branches of solution; \citealt{Shut2001}). This is in contrast with the non-rotating scenario, best thought of as the limit $f \to 0$ of the general situation. In this limit, the inertial levels coalesce with the critical level, leading to a stronger, $k$-independent singularity and to a singular behaviour of the wavepacket unless dissipation is introduced. 
(A similar smearing out of singularities across different altitudes  also occurs without rotation when more complicated flows, such as the directional shear flow, are considered; \citealt{Shut1995,Shut2003}, \citealt{Mart2007}.)

The term `inertial layer' is used to describe the region where the effect of the inertial-level singularities is distributed. It is centred around the dominant inertial level, 
\begin{equation}
\tz=-\frac{f}{k_*\Lambda},
\end{equation}
determined  by the central wavenumber $k_*$ of the topography. Note that, for the problem under consideration, only the lower inertial levels matter since the waves are exponentially small in $J$ at the upper inertial levels. Condition (\ref{prop}) ensures that $z_* > -H$, that is, the dominant inertial level lies in the fluid domain.
The characteristic thickness of the inertial layer is found as 
\begin{eqnarray}
\delta_* = -\frac{\tz}{\tk}\delta k = \frac{f}{\tk^2 \Lambda \Delta}, \label{ilscale}
\end{eqnarray}
on using that, according to expression  (\ref{mountain}) for the mountain height, the spectral width of the wavepacket is $\delta k = \Delta^{-1}$. 

%where we have taken into account that in the topography expression (\ref{mountain}) the wavepacket width in spectral space $\delta k = O(\Delta^{-1})$, and 
%denotes the dominant inertial level where the wave with single dominant wavenumber $\tk$ reaches its inertial level. 
%In Section \ref{Sec_wm}, we will show that for a uniform-shear background flow, in contrast to the no rotation situation, wave force on the mean flow is not singular in the finite-thickness inertial layer when rotation is considered.
%In this case the divergence of EP flux hence the wave force on the mean flow is no longer singular because the force region is expanded to a inertial layer.
%Hence, we pay special attention to the inertial layer because this is the region responsible for wave-mean flow interaction, whose details are shown in the Section \ref{Sec_wm}.

Equation (\ref{WaveEq}) for $\hat w$ can be solved explicitly in terms of hypergeometric functions \citep{Yama1984,Shut2001}.
We rely instead on the approximate WKB solution derived in \citet{Lott2012} (see also \citet{Lott2015}) and valid in the large-Richardson limit $J \gg 1$.  
In {this solution, $\hat w$ is approximated as}
\begin{equation}
\hat {w}{(k,l,z)}=\frac{\zeta}{\zeta_{\mathrm{b}}}\br{\frac{\zeta_{\mathrm{b}}-1}{\zeta-1}}^{1/4-\ii \nu/2}
\br{\frac{\zeta_{\mathrm{b}}+1}{\zeta+1}}^{1/4+\ii \nu/2} 
\ex^{-\ii J \sqrt{1+\nu^2} D(\zeta)},\label{vertial_str0}
\end{equation}
where
\begin{equation}
D(\zeta) = \ln(\zeta+\sqrt{\zeta^2-1}) - \ln(\zeta_{\mathrm{b}}+\sqrt{\zeta_{\mathrm{b}}^2-1}), \quad \textrm{with} \ \ 
\zeta_\mathrm{b}=k\Lambda H/f.
\end{equation} 
Note that we have normalized  $\hat w$ so that its bottom-boundary value is $\hat w (\zeta_\mathrm{b})=1$.

{
Expression (\ref{vertial_str0}) holds for all real values of $\zeta$ except in small regions of $O(J^{-2})$ thickness around the inertial levels $\zeta = \pm 1$. The sign of the argument of the exponential is taken to be negative because this ensures that the wave is propagating upwards above the upper inertial level, that is, for $\zeta < -1$ (see \citet{Book1967}; note that the opposite, positive sign is found when the background velocity is increasing with altitude). The fractional powers and logarithms involved in (\ref{vertial_str0}) are multivalued functions for which suitable branches need to be selected. This selection is dictated by causality and is most easily settled by adding small damping terms in (\ref{WavePri}). The upshot is that the multivalued functions should be continued from $\zeta<-1$ to $\zeta>1$ along a contour in the complex plane that passes \textit{below} the singularities at $\zeta=\pm1$ \citep{Book1967,Jones1967,Lott2015}. In this way, $\sqrt{\zeta^2-1}=-\ii \sqrt{1-\zeta^2}$ between the inertial levels so that the solution is decreasing exponentially with $\zeta$  there, like $\exp(-J\sqrt{1+\nu^2} \cos^{-1} \zeta)$. Overall, the solution experiences an absorption by the factor $\exp(-J\sqrt{1+\nu^2} \pi)$ (see also \cite{Lott2012}) known to apply to both the rotating and non-rotating cases. Note that the reliance on causality means that the problem is not treated as strictly inviscid but rather as a vanishing viscosity limit.
}

%Expression (\ref{vertial_str0}) holds for all real values of $\zeta$ except in small regions of $O(J^{-2})$ thickness around the inertial levels $\zeta = \pm 1$, provided that suitable branches of the multivalued functions are chosen (see \citet{Jones1967,Lott2012,Lott2015}). The sign of the argument of the exponential is taken to be negative; this ensures that the wave is propagating upwards above the upper inertial level, that is, for $\zeta < -1$ (see \citet{Book1967}; note that the opposite, positive sign is found when the background velocity is increasing with altitude). The multivalued functions are continued from $\zeta>1$ to $\zeta<-1$ along a contour in the complex plane that passes below the singularities at $\zeta=\pm1$.
%In this way, the solution is decreasing exponentially with $\zeta$ between the inertial level (like $\exp(-J\sqrt{1+\nu^2} \cos^{-1} \zeta)$) and experiences an overall absorption by the factor $\exp(-J\sqrt{1+\nu^2} \pi)$ known to apply to both the rotating and non-rotating cases \citep{Book1967,Jones1967,Lott2015}. The choice of continuation is dictated by considerations of causality which are readily settled by adding small damping terms in (\ref{WavePri}). 

The WKB approximation (\ref{vertial_str0}) breaks down near the inertial levels, specifically for $||\zeta|-1| = O(J^{-2})$ where it should be replaced by an expression in terms of Hankel functions \citep{Lott2012}. These regions are narrow enough and the singularities of (\ref{vertial_str0}) at $\zeta=\pm1$ are mild enough  that they can be ignored when computing the vertical velocity of the complete wavepacket.
%\commentJV{is this strictly true, or are the regions negligible only as far as the mean-flow forcing is concerned? In any case, where is this shown below?}

%Even though we consider a steady wave, we need to choose the branch corresponding to upward group velocity -- in the exponential of (\ref{vertial_str0}) the minus sign is chosen instead of a plus -- to be physically sensible in the perspective of steady wave generation. 
%This branch choice is a consequence of the assumption that the far-field wave condition above the singularity only contains an upward propagating wave. 
%In fact we can choose different far-field conditions or boundary conditions to include wave reflection, but because the wave field is linear it will only digress us from our aim of understanding the wave and mean flow behaviour around the singularities.
%In the vicinity of $\zeta = 1$, specifically $|\zeta-1|=O(J^{-2})$, $\hat{w}$ is expressed by Hankel functions.
%And when $\zeta<1$, $\hat{w}$ is exponentially small.

Substituting the form (\ref{mountain}) of the topography into the boundary condition (\ref{w_bc}), we obtain the 
vertical velocity at the boundary as the Fourier expansion
\begin{equation}
w_{1\mathrm{b}} = \frac{\ii U_{\mathrm{b}}h\Delta^2}{2\pi}\int_{-\infty}^{\infty} \int_{-\infty}^{\infty}  k \ex^{-|\bk-\bk_*|^2\Delta^2/2}\ex^{\ii \bk \cdot \bx} \, \dd k \dd l. \label{w1b}
\end{equation}
Here and henceforth, a real part is implied. %\commentJV{The real part is important: have you taken it into account when working out the wave forcing?}
Combining this with (\ref{hatwdef}) and  (\ref{vertial_str0}) leads to the vertical velocity of the wavepacket in the form
\begin{equation}
w_1 = \frac{\ii U_{\mathrm{b}} h \Delta^2}{2\pi}\int_{-\infty}^{\infty} \int_{-\infty}^{\infty} 
k \ex^{-|\bk-\bk_*|^2\Delta^2/2}
\hat{w} \, \ex^{\ii \bk \cdot \bx} \, \dd k \dd l. \label{WOri0}
\end{equation}
Note that, because $\hat w$ is exponentially small for $\zeta < 1$ as a result of wave absorption, the lower limit of the integral in $k$ could be taken as  $-f\Lambda^{-1}z^{-1}$. This absorption is crucial for the impact of the wavepacket on the mean flow.

%Here, from the definition of inertial level -- $\zeta=1$ -- the integration limit $(-f\Lambda^{-1}z^{-1},\infty)$ captures the wavenumbers whose inertial level above $z$ with an omission of high order terms. 
%The exponentially small $\zeta<0$ contribution is directly neglected, while the $|\zeta-1|=O(J^{-2})$ contribution is neglected because the region is of high order smaller than the effective integration region which we will verify later, and the Hankel function does not contain singularity in this region.
%This wavenumber truncation approximates the branch choice with the minus sign in the exponential of (\ref{vertial_str0}) across the inertial level when the terms in the square root become negative, so with an exponentially small error the lower integration limit can be replaced by $-\infty$.
%This phenomenon of wave absorption across the singularity further results in EP flux loss, which is crucial for topographic wave--shear mean flow interaction. 

\subsection{Saddle-point approximations}
\label{sec:saddles}

The solution (\ref{WOri0}) can be further simplified by taking advantage of the assumptions $J\gg 1$ and $k_*\Delta \gg 1 $ to apply a saddle-point approximation.
The key is  to identify the dominant terms in the argument of the exponential, including a contribution from $\hat w$.
To avoid defining several new dimensionless numbers measuring the relative size of $J$, $k_*\Delta$ and $\Ro$, it is expedient to introduce a bookkeeping parameter $\epsilon$ which keeps track of the orders of various terms.
This parameter is treated as formally small and used as a basis for a saddle-point approximation, but it is set to $1$ at the end of the computation to obtain asymptotic formulas in a convenient dimensional form.
%Here we need to emphasize that this bookkeeping parameter is different from the nondimensional parameters, where the former is simply a mark to remind the orders of different terms while the latter are real parameters controlling the system.
The bookkeeping parameter $\epsilon$ is introduced through the replacements
\begin{equation}
{\bk}_*\Delta \mapsto \epsilon^{-1} {\bk}_*\Delta, \quad J \mapsto \epsilon^{-1}J \quad \mathrm{and} \quad \Ro \mapsto \epsilon \Ro, 
\label{eps}
\end{equation}
in accordance with the {scalings (\ref{smallRossby}), (\ref{Jscale}) and (\ref{kDscale})}. Thus the formal smallness of $\epsilon$ captures at once the mountain scale separation, the large Richardson number, and the small Rossby number. {We emphasise that the bookkeeping device is completely equivalent to using, say, $J$ as a large parameter and treating $\bk_* \Delta/J$ and $J \Ro$ as order one; it is employed here for the economy and transparency of notation it brings.}

%\begin{subequations}
%\begin{align}
%w_{1\mathrm{b}} &= \frac{U_{\mathrm{b}}h\Delta^2}{2\pi}\int_{-\infty}^{\infty} \int_{-\infty}^{\infty} \ii k \ex^{-|\bk-\bk_*|^2\Delta^2/2}\ex^{\ii \bk \cdot \bx} \dd k \dd l \\
%&= \frac{\ii U_{\mathrm{b}} h k_*\Delta^2}{2\pi}\int_{-\infty}^{\infty} \int_{-\infty}^{\infty} \ex^{-|\bk-\bk_*|^2\Delta^2/2}\ex^{\ii \bk \cdot \bx} \dd k \dd l, 
%\end{align}
%\end{subequations}
%where in the second step 

Introducing (\ref{eps}) into (\ref{WOri0}) leads to
\begin{eqnarray}
w_1 = \frac{\ii U_{\mathrm{b}} h k_*\Delta^2}{2\pi\epsilon}\int_{-\infty}^{\infty} \int_{-\infty}^{\infty} 
\ex^{-|\bk-\epsilon^{-1}\bk_*|^2\Delta^2/2}
\hat{w} \ex^{\ii \bk \cdot \bx} \,  \dd k \dd l, \label{WOri}
\end{eqnarray}
where 
\begin{equation}
\hat{w}=\frac{\zeta}{\zeta_{\mathrm{b}}}\br{\frac{\zeta_{\mathrm{b}}-1}{\zeta-1}}^{1/4-\ii \nu/2}
\br{\frac{\zeta_{\mathrm{b}}+1}{\zeta+1}}^{1/4+\ii \nu/2} 
\ex^{-\ii \epsilon^{-1} J \sqrt{1+\nu^2} D(\zeta)}, \label{vertial_str}
\end{equation}
and makes the dependence on $\epsilon$ explicit. In writing this expression, we have made a first approximation
by replacing the wavenumber $k$ outside the exponential functions by its leading-order approximation $k_*$. The error introduced is negligible as can be verified below once the size of the neighbourhood of $k_*$ controlling the integral is estimated.

A second approximation is made in carrying out the integration with respect to $l$ in (\ref{WOri}). Because the integral is dominated by values of $l$ near $\epsilon^{-1} l_*$, we write
\begin{eqnarray}
l = \epsilon^{-1}l_*+L, \label{expan_l}
\end{eqnarray}
where $\epsilon L/l_* \ll 1$, leading to the following expressions
\begin{eqnarray}
\nu &=& \frac{l_*}{\epsilon k}\br{1+\epsilon\frac{L}{l_* }},\\
\sqrt{1+\nu^2} &=& \sqrt{1+\br{\frac{l_*}{\epsilon k}}^2}+ \epsilon \frac{l_* L}{\epsilon k\sqrt{(\epsilon k)^2+l_*^2}} + O\br{\br{\epsilon\frac{L}{l_*}}^2}, \label{lexpan}
\end{eqnarray}
where $\epsilon k$ is treated as $O(1)$ since $k$ is close to $\epsilon^{-1}k_*$.
Substituting these into (\ref{WOri}) and neglecting $O((\epsilon L/l_*)^2)$,  we obtain
\begin{equation}
\begin{aligned}
w_1&= \frac{\ii U_{\mathrm{b}}hk_*\Delta^2}{2\pi\epsilon} \int_{-\infty}^{\infty}\int_{-\infty}^{\infty}
\ex^{-(\Delta L)^2/2 - \ii Jl_* (\epsilon k)^{-1} ((\epsilon k)^2+l_*^2)^{-1/2} D(\zeta)L + \ii L y} \dd L  \\
&\qquad \qquad \qquad  \qquad \qquad 
\times\ex^{-(k-\epsilon^{-1}k_*)^2\Delta^2/2} \frac{\zeta}{\zeta_{\mathrm{b}}} 
\br{\frac{\zeta_{\mathrm{b}}-1}{\zeta-1}}^{1/4-\ii\nu_*/2} \br{\frac{\zeta_{\mathrm{b}}+1}{\zeta+1}}^{1/4+\ii\nu_*/2} \\
&\qquad  \qquad  \qquad  \qquad \qquad  \qquad   \times
\ex^{-\ii J \epsilon^{-1}(1+l_*^2/(\epsilon k)^2)^{1/2} D(\zeta)+ \ii k x }\, \dd k  \\
&\doteq \frac{U_{\mathrm{b}}hk_*\Delta}{\epsilon\sqrt{2\pi}} \int_{-\infty}^{\infty} 
\ex^{-\br{ y - Jl_* (\epsilon k)^{-1} ((\epsilon k)^2+l_*^2)^{-1/2}D(\zeta) }^2/(2 \Delta^2)} \\
&\qquad  \qquad  \qquad   \qquad  \qquad  \times\ex^{-(k-\epsilon^{-1}k_*)^2\Delta^2/2} \frac{\zeta}{\zeta_{\mathrm{b}}}
\br{\frac{\zeta_{\mathrm{b}}-1}{\zeta-1}}^{1/4-\ii\nu_*/2}\! \br{\frac{\zeta_{\mathrm{b}}+1}{\zeta+1}}^{1/4+\ii\nu_*/2} \\
&\qquad  \qquad  \qquad  \qquad \qquad  \qquad  \times
\ex^{-\ii J \epsilon^{-1}(1+l_*^2/(\epsilon k)^2)^{1/2} D(\zeta)+ \ii k x } \, \dd k. \label{w}
\end{aligned}
\end{equation}
In the second line, we have ignored a phase factor and introduced the symbol $\doteq$ to denote an equality in modulus only, ignoring phase factors. In what follows, we pay only attention to the modulus of $w_1$ since this controls the wave--mean flow interaction properties: the spatially averaged EP flux, which is quadratic in wave quantities, only depends on the wave amplitude and on the relative phase of various fields which is easily worked out. 
%\commentJV{This is just not true: the phase of $w_1$ is needed to get $w_{1z}$
%.} \commentJV{I have edited the formula to avoid fractions in the exponentials which lead to very small fonts; I also got rid of the prefactor $\ii$ which is also a phase factor.}

The appearance of the bookkeeping parameter $\epsilon^{-1}$ in the exponential in  (\ref{w}) motivates the saddle-point approximation. To apply this, we need to compare the leading terms in the exponential, namely
\begin{equation}
(k-\epsilon^{-1}k_*)^2\Delta^2/2 \quad \text{and} \quad \ii J \epsilon^{-1}\left(1+\br{\frac{l_*}{\epsilon k}}^2\right)^{1/2} D(\zeta). \label{expbal}
\end{equation}
The first term stems from the finite spectral width of the mountain height; the second term, which depends on $\zeta$ and hence $z$, captures the vertical structure of the wave. We seek distinguished regimes, where the leading-order terms in (\ref{expbal}) balance. This requires approximating $D(\zeta)$ to determine its order, a non-trivial task since the order of $D(\zeta)$ depends on the value of $\zeta$, that is, on the particular range of altitude considered.
Mathematically, different altitude ranges are captured by different values of $\alpha$ in the scaling
\begin{equation}
z = z_* \br{ 1 + \br{\frac{\epsilon}{k_*\Delta}}^{\alpha} Z }, \label{z_exp}
\end{equation}
where $Z=O(1)$.
% and, because of the introduction of bookkeeping parameter, $z_*=-\epsilon f/(k_*\Lambda)$. \commentJV{Not sure about this last point: there's essentially a rescaling $\Lambda \to \epsilon \Lambda$ involved.}
%Larger values of $\alpha$ correspond to regions of $z$ closer the dominant inertial level. 

Note that, since $z_*<0$, $Z$ and $z$ have different signs, with $Z>0$ below the dominant inertial level and increasing downwards. 
Because $\zeta$ depends on both $z$ and $k$, we also need to scale $k$ to find  range of wavenumbers controlling the integral in (\ref{w}). Since the wavepacket is concentrated around $\epsilon^{-1} k_*$, we write 
\begin{equation}
k=\epsilon^{-1}k_*\br{1+\br{\frac{\epsilon}{k_*\Delta}}^{\beta}K}, \quad \mathrm{with} \quad K=O(1) \quad \mathrm{and} \quad \beta\geq 0. \label{k_exp}
\end{equation}

Combining  (\ref{z_exp}) and (\ref{k_exp}), we obtain
\begin{equation}
\zeta = -\frac{k\Lambda z}{f} = 1+\br{\frac{\epsilon}{k_*\Delta}}^{\beta}K + \br{\frac{\epsilon}{k_*\Delta}}^{\alpha} Z + O\br{\br{\frac{\epsilon}{k_*\Delta}}^{\alpha + \beta}}.
\end{equation}
We now need to distinguish two situations: 
(i) for $\alpha=0$, that is, away from the dominant inertial level, 
\begin{equation}
D(\zeta) = D(1+Z) + O\br{\br{\frac{\epsilon}{k_*\Delta}}^{\beta}}; \label{d_exp2}
\end{equation}
(ii) for $\alpha>0$, that is, asymptotically close to the dominant inertial level,
\begin{equation}
D(\zeta) = D(1) + O\br{\br{\frac{\epsilon}{k_*\Delta}}^{\min\{\alpha,\beta\}/2}}. \label{d_exp}
%= - \ln(\zeta_{\mathrm{b}}+\sqrt{\zeta_{\mathrm{b}}^2-1}) +  O\br{\br{\frac{\epsilon}{k_*\Delta}}^{\min\{\alpha,\beta\}/2}}. 
\end{equation}
Because the leading-order terms in these two expansions are independent of $k$, hence do not contribute to the integration over $k$ in (\ref{w}), the order of the second terms is crucial. Using the scalings of the second terms of (\ref{d_exp2})--(\ref{d_exp}) in the second expression in (\ref{expbal}), and balancing with the first expression (scaling like $\epsilon^{2 \beta -2}$) leads to two distinguished regimes: Regime I, with $\alpha = 0$ and $\beta=1$, and Regime II, with $\alpha = \beta = 2/3$ (since $\alpha=\beta$ gives a distinguished regime). In each regime, the coordinate $x$ appearing in $\exp(\ii k x)$ should be scaled so that the $K$-dependent contribution to $k x$, proportional to $\epsilon^{\beta-1}$, be of the same order as the $K$-dependent terms in (\ref{expbal}). This leads to $x=O(\Delta)$ in Regime I and $x=O(\epsilon^{-1/3}\Delta)$ in Regime II. 

We carry out the saddle-point expansion of (\ref{w}) in these two regimes in Appendix \ref{appendix1} 
and only quote the final results here. In Regime I, after setting $\epsilon=1$, we  find that
\begin{equation}
w_1 \doteq h f (1+Z_{\mathrm{I}}) \br{\frac{r^2-1}{2Z_{\mathrm{I}}+Z_{\mathrm{I}}^2}}^{1/4}
\ex^{-\br{ y - J \nu_*{ k_*}^{-1} (1+\nu_*^2)^{-1/2}D_\mathrm{I} }^2/(2 \Delta^2)}
\ex^{-{X_{\mathrm{I}}^2}/(2\Delta^2)}, \label{approx_I}
\end{equation}
where 
\begin{equation}
X_{\mathrm{I}} = x -\frac{J}{k_*}\left(
\sqrt{1+\nu_*^2}
\br{ 
	\frac{1+Z_{\mathrm{I}}}{\sqrt{2Z_{\mathrm{I}}+Z_{\mathrm{I}}^2}} - \frac{r}{\sqrt{r^2-1}} }
-\frac{\nu_*^2}{\sqrt{1+\nu_*^2}}D_\mathrm{I}\right),
\end{equation}
$D_\mathrm{I}=\ln(1+Z_{\mathrm{I}}+\sqrt{2Z_{\mathrm{I}}+Z_{\mathrm{I}}^2}) - \ln(r+\sqrt{r^2-1})$, $Z_\mathrm{I} = z/{z_*}-1$ and {we have used that $r=k_*\Lambda H/f$}.
This makes clear that the wavepacket retains the bell shape of topography, with scale $\Delta$, throughout its propagation across Regime I. The path of the wavepacket in the $(x,z)$-plane is determined by setting $X_\mathrm{I}=0$ in (\ref{approx_I}), which corresponds to standard ray tracing.
%\commentJV{You wrote that the Appendix shows this is ray tracing, but I couldn't find anything apart from a repeat of the claim.}
It shows in particular that the wavepacket diverges to infinity as it approaches the dominant inertial level, with $x \sim J k_*^{-1} (1+\nu_*^2)^{1/2} (2 Z_\mathrm{I})^{-1/2}$ as $Z_\mathrm{I} \to 0$. This is a limitation of the approximation made in Regime I (also a limitation of ray tracing) rather than a physical effect as the analysis of Regime II shows.
%
%
%In addition, in the Appendix \ref{RegimeI} we can see that the derivation of (\ref{approx_I}) is in fact the derivation of ray tracing, which indicates the wavepacket's scales.
%Note that the ray tracing result suggests that the wavepacket tends to $x\to\infty$ as $z$ approaches $z_*$.
%In our solution, we apply an implicit dissipation corresponding to the branch choice \cite{Book1967} in the solution of (\ref{WaveEq}) to obtain a physically reasonable result.

In Regime II, and again with $\epsilon=1$, we find that
\begin{equation}
w_1 \doteq \frac{ U_{\mathrm{b}}hk_*^2\Delta}{2^{1/4}(k_*\Delta)^{3/4}} \frac{q(K_\mathrm{s})}{(p''(K_\mathrm{s}))^{1/2}} \,
\ex^{(k_*\Delta)^{2/3}p(K_\mathrm{s})} g(y), \label{ApprII}
\end{equation}
where the $y$-dependence is controlled by the Gaussian
\begin{equation}
g(y) = \ex^{-\br{ y - J \nu_*{ k_*}^{-1} (1+\nu_*^2)^{-1/2}D_\mathrm{II} }^2/(2 \Delta^2)}, 
\label{g(y)}
\end{equation}
with $D_{\mathrm{II}} = -\ln(r+\sqrt{r^2-1})$,
and is decoupled from the dependence in $x$ and $z$. 
In (\ref{ApprII}), the function $p$ is defined by
\begin{equation}
p(K_{\mathrm{s}}) = -\frac{K_{\mathrm{s}}^2}{2} - \ii J(k_*\Delta)\sqrt{1+\nu_*^2}\sqrt{2(K_{\mathrm{s}}+Z_{\mathrm{II}})} + \ii K_{\mathrm{s}} X_{\mathrm{II}},
\label{phasep}
\end{equation}
and $K_\mathrm{s}$ is one of its saddle points, satisfying $p'(K_\mathrm{s})=0$, where the prime denotes derivative.
The other symbols introduced are
\begin{equation}
q(K_{\mathrm{s}}) = \frac{1}{(K_{\mathrm{s}}+Z_{\mathrm{II}})^{1/4}},
\end{equation}
\begin{equation}
X_{\mathrm{II}}= (k_*\Delta)^{-4/3}\br{k_* x + J\left(
	\frac{r \sqrt{1+\nu_*^2}}{\sqrt{r^2-1}} + 
	\frac{\nu_*^2}{\sqrt{1+\nu_*^2}}D_\mathrm{II}
	\right) },
\end{equation}
and $Z_\mathrm{II}=(k_*\Delta)^{2/3}(z/z_*-1)$. The saddle point $K_\mathrm{s}$ satisfies a cubic equation whose analytic solution is not particularly illuminating; we will solve it numerically. It is selected among the three roots of the cubic by the condition that it be accessible by a steepest-descent path connecting $-\infty$ to $\infty$ {\citep[e.g.][]{Bend1999}}.
For large $Z_{\mathrm{II}}$, the expansion $\sqrt{2(K_\mathrm{s} + Z_{\mathrm{II}})} \sim \sqrt{2 Z_{\mathrm{II}}} + K_\mathrm{s}/\sqrt{2 Z_{\mathrm{II}}}$ can be used to confirm the matching between Regime I and Regime II. 
%\commentJV{I haven't fully checked this, but the phase functions seem to tend to one another.}
%Here the ray tracing is no longer valid when the wavepacket is close to the inertial layer, and the wavepacket is absorbed at a finite horizontal position but with a rescaled horizontal structure, moreover, the wave amplitude is finite. 
The behavior for small $Z_\mathrm{II}$ is key for the mean-flow forcing. As discussed in \S\ref{sec_wave_pre}, the change in EP flux is concentrated in the inertial layer, such that $|z-z_*|=O(\delta_*)$ with $\delta_*$ given in (\ref{ilscale}). This corresponds to the scaling
\begin{equation}
Z= \epsilon^{-1} {k_*\Delta}\br{\frac{z}{z_*}-1} = O(1),
\label{Z}
\end{equation}
hence to $\alpha=1$ in (\ref{z_exp}) and thus $Z_\mathrm{II}=O(\epsilon^{1/3}) \ll 1$. For this range of $Z$, the integral in (\ref{w}) is dominated by wavenumbers $k$ in an $O(1)$ neighbourhood of the central wavenumber $\epsilon^{-1} k_*$, corresponding to $\beta=1$ in (\ref{k_exp}). The associated regime, which we term Regime II$_\mathrm{B}$, is a limit of Regime II, obtained when some terms are negligible (notably the first term in the phase function (\ref{phasep})). 
This makes it possible to derive an expression for $w_1$ simpler than (\ref{ApprII})which we will subsequently use to compute the mean-flow forcing. This expression is derived in Appendix \ref{RegimeIIB} and given by
%\begin{equation}
%w_1 \doteq   \frac{h f J^{1/2}\br{1+\nu_*^2}^{1/4} \br{r^2 -1}^{1/4}}{(k_*\Delta)^{1/2} X} \,
%\ex^{ -\left(J^2\br{1+\nu_*^2} /(2 (k_* \Delta X)^{2})-Z\right)^2/2} g(y), \label{wB}
%\end{equation}
\begin{equation}
w_1 \doteq   \frac{h f J^{1/2}\br{1+\nu_*^2}^{1/4} \br{r^2 -1}^{1/4}}{(k_*\Delta)^{1/2} X} \,
\ex^{ -\left(a^2/X^{2}-Z\right)^2/2} g(y), \label{wB}
\end{equation}
where
\begin{equation}
a = \frac{J \sqrt{1+\nu_*^2}}{\sqrt{2} k_* \Delta}
\label{a}
\end{equation}
and
\begin{equation}
X= (k_*\Delta)^{-3/2}\br{k_* x + J\left(
	\frac{r \sqrt{1+\nu_*^2}}{\sqrt{r^2-1}} + 
	\frac{\nu_*^2}{\sqrt{1+\nu_*^2}}D_\mathrm{II}
	\right) }. \label{Xrescale_IIB}
\end{equation}
%\commentJV{I have changed the sign of the second term (multiplying $J$) in the definition of $X$; can you check this is correct?}

According to this, the wavepacket is centred on the curve
\begin{equation}
X = X_{\mathrm{c}}=\br{ \frac{J^2\br{1+\nu_*^2}}{(k_* \Delta)^2(Z+\sqrt{2+Z^2})} }^{1/2} \label{peakIIB}
\end{equation}
in the $(X,Z)$-plane (obtained by maximising (\ref{wB})) and localized in a region of order-one size in both the $X$ and $Z$ direction. In view of (\ref{Z}) and (\ref{Xrescale_IIB}), this corresponds to a region of streamwise extent $O((k_*\Delta)^{1/2}\Delta)$, thus much larger than the size $\Delta$ of the mountain, located an $O((k_*\Delta)^{1/2}\Delta)$ distance downstream of the topography, and of $O(\delta_*)$ vertical extent.  
Thus, the prediction of ray tracing of a wavepacket that diverges to infinity as the dominant inertial level is approached is replaced in Regime II$_\mathrm{B}$ (and hence in Regime II) by a large-but-finite horizontal shift. 
{Note that in the limit of $f\to 0$, the wavepacket propagates vertically with a vertical velocity that tends to zero as it approaches the critical level $z = 0$ towards which the inertial levels coalesce, with $w_1 \propto z^{1/2}$ as $z \to 0$. This can be deduced from (\ref{approx_I}) 
by letting $r \to \infty$ and is consistent with a direct computation assuming $f=0$. For a small-but-finite $f$, however, there is an extremely thin inertial layer of size $O((k_* \Delta)^{-2} \Ro^{-1} H)$, in which the wavepacket experiences the horizontal displacement described by (\ref{Xrescale_IIB}).}

We remark that the three regimes identified by the saddle-point analysis can be interpreted physically. Regime I is the ray-tracing regime, which is unaffected by the singularities of the wave solution. Regime II is controlled by the singularity at the lower inertial level, and Regime II$_\mathrm{B}$ is its part dominated by inertial-level absorption. 
%The valve effect, that is, the distinct behaviours of waves with $l$ positive and negative between the inertial levels 
%\commentJV{Do we want to say something about the shift in $y$ of the wavepacket? Or about the valve effect which seems to play no role.}

%in Regime II the singular effect due to the vertical structure function of a single-frequency wave starts to influence the wave packet;
%in Regime II$_\mathrm{B}$, another singular effect -- inertial level absorption -- is dominant, and the region width calculated from the definition of inertial level is controlled by the spectrum width of the mountain. 
%In these regions when the smallest integral domain of $k$ is taken in Regime II$_\mathrm{B}$ the range of $\zeta$ is $|\zeta-1|=O(J^{-1/2})$ which is much larger than the region with $\hat{w}$ in the form of Hankel function approximation, so the approximation of the integral domain truncation in (\ref{WOri0}) is verified.

\subsection{Numerical results}

In this section, we compare the asymptotic predictions for $w_1$ with direct numerical computations of the integral in (\ref{w}). We first take the parameters 
\begin{equation}
\Ro = 0.02, \ \ k_*\Delta=100, \ \ l_*\Delta=100, \ \ J=100 \ \ \textrm{and} \ \ \nu_*=1,
\label{param1}
\end{equation} 
so $r =k_*\Delta \Ro =2$. {A choice of physical parameters leading to these values is 
$N=1.4\times 10^{-2}\,\mr{s}^{-1}$, $f=10^{-4}\,\mr{s}^{-1}$, $\Lambda=1.4\times 10^{-4}\,\mr{s}^{-1}$, $H=5\,\mr{km}$, $\Delta=3.5\times 10^2\,\mr{km}$ and $k_*=2.8\times 10^{-4}\mr{m}^{-1}$. The weak shear and short wavelengths make this choice somewhat contrived, but it has the advantage of enabling a comparison in conditions where the asymptotic assumptions hold unambiguously. Parameters corresponding to a stronger shear and longer wavelengths are considered at the end of the section. } 
We concentrate on the amplitude $|w_1|$ in the $(x,z)$ cross-section where it is maximum, since the structure in $y$ is simply the Gaussian structure of the mountain envelope, albeit with a shift.

\begin{figure}
	\centering
	\includegraphics[width=0.9\textwidth]{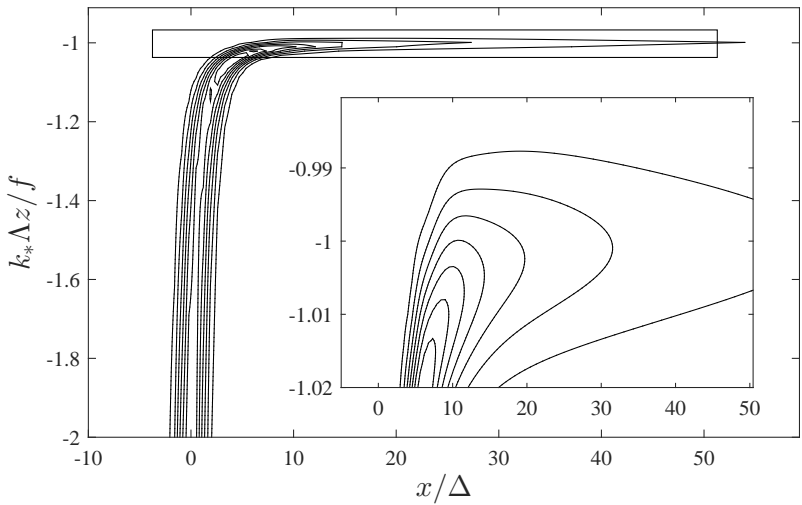}
	\caption{Vertical velocity amplitude $|w_1|$  in the $(x,z)$ plane obtained by numerical integration of (\ref{WOri}) for the parameters in (\ref{param1})  (7 equispaced contours with velocity in the range $[0.04,0.28] \times Uhk_*/\sqrt{2\pi}$  are shown). The inset is a zoom on the rectangle region indicated in the main panel.}\label{FigContour}
\end{figure}

Figure \ref{FigContour} shows a contour plot of $|w_1|$ obtained numerically.
Away from the dominant inertial level, $k_*\Lambda z/f = -1$, the wavepacket is only sightly deflected from the vertical. Closer to the dominant inertial level, the wavepacket shows a significant bend; we can read off from the inset that the peak of the wavepacket at the dominant inertial level is around $x/\Delta \approx 12$, in agreement with the peak value predicted in (\ref{Xrescale_IIB}) which gives $X_{\mathrm{c}}=2^{1/4}$.
This figure is more a qualitative illustration than a quantitative comparison which we carry out next.

\begin{figure}
	\centering
	\includegraphics[width=0.95\textwidth]{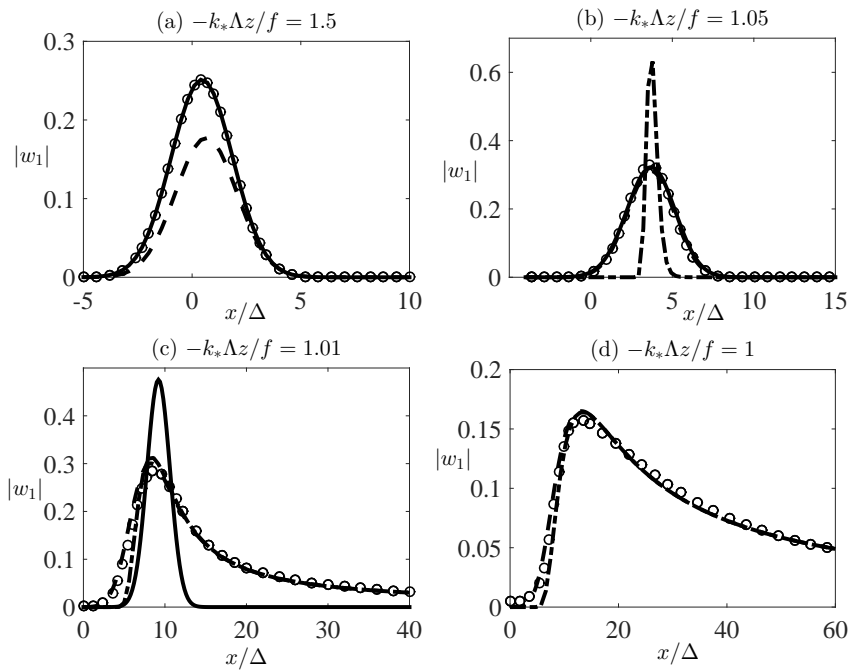}
	\caption{Vertical velocity amplitude $|w_1|$ normalized by $Uhk_*/\sqrt{2\pi}$ as a function of $x/\Delta$ for the altitudes given by
		$-k_*\Lambda z/f=1.5$ (a), 1.05 (b), 1.01 (c) and 1 (d), and the parameters in (\ref{param1}). 
		Numerical results (circles) are compared with the asymptotic predictions of Regime I (solid line, shown in panels (a)--(c)), Regime II (dashed line), and Regime II$_\mathrm{B}$ (dash-dotted line, shown in panels (b)--(d)). 
	}\label{FigWave}
\end{figure}

Figure \ref{FigWave} provides a detailed comparison of the numerical and asymptotic predictions for $|w_1|$. Its four panels show $|w_1|$ as a function of $x/\Delta$ at four different altitudes. 
Panel (a) corresponds to $-k_*\Lambda z/f=1.5$, sufficiently far below the dominant inertial level for the Regime I asymptotics to apply. As expected, the asymptotic predictions of Regime I (solid line) matches the numerical results (circles), with a wavepacket that takes the bell shape of the mountain enevelope, while the predictions of Regime II (dashed line) do not. 
Panel (b) shows the wavepacket closer to the dominant inertial level, for $-k_*\Lambda z/f=1.05$, in a region where Regime I and Regime II overlap: the predictions of both regimes match the numerical results. 
The Regime II$_{\mathrm{B}}$ approximation is also shown (dash-dotted line) and, unsurprisingly, is found to be invalid.  
%From both of the two figures in the upper row we can observe the slight horizontal bending of the wavepacket. 
Closer still to the dominant inertial level, as shown in panel (c) for $-k_*\Lambda z/f=1.01$, the Regime I approximation breaks down. The Regime II predictions match the numerical results closely, while those of Regime II$_\mathrm{B}$ are accurate for $x$ large enough. As expected, the wavepacket is no  longer bell shaped, and its peak is shifted by an $O((k_*\Delta)^{1/2}\Delta)$ amount to the right (since $x/\Delta \approx 10 = (k_*\Delta)^{1/2}$) in agreement with (\ref{Xrescale_IIB}) and (\ref{peakIIB}). 
Finally, at the dominant inertial level as shown in panel (d), the predictions of both Regime II and Regime II$_\mathrm{B}$ coincide and match the numerical results in most of range of $x$, except to the very left of the peak where Regime II$_\mathrm{B}$ underestimates the amplitude. As discussed above, the peak of the wavepacket remains at a finite $O((k_*\Delta)^{1/2}\Delta)$ position. Crucially for mean-flow forcing, the maximum amplitude is also strongly reduced as a result of absorption.

\begin{figure}
	\centering
	\includegraphics[width=0.95\textwidth]{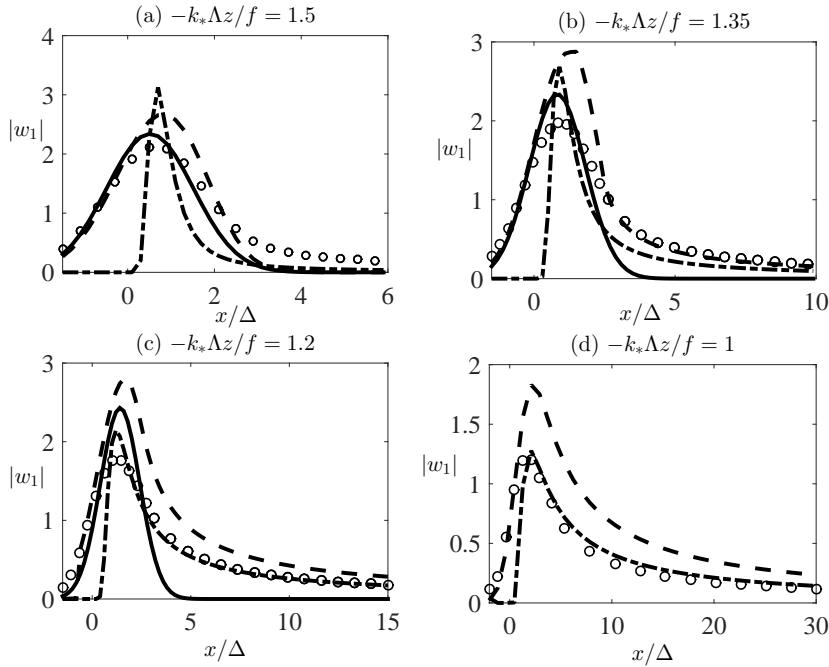}
	\caption{Same as Fig.\ \ref{FigWave} but for the parameters in (\ref{param2}) and for
		$-k_*\Lambda z/f=1.5$ (a), 1.35 (b), 1.2 (c), 1 (d). %\commentJV{Labels should be $|w_1|$. Number of circles could be reduced, except in panel (d) where it's good (but are the circles not equidistant in $x$? There's a gap for $x \approx 5 \Delta$.} 
	}\label{FigWave2}
\end{figure}

We now consider a more realistic parameter choice relevant to the atmosphere. 
{
Taking $N=1.4\times 10^{-2}\,\mr{s}^{-1}$, $f=10^{-4}\,\mr{s}^{-1}$, $\Lambda=5.6\times 10^{-3}\,\mr{s}^{-1}$, $H=5\,\mr{km}$, $\Delta=3.5\times 10^2\,\mr{km}$ and $k_*=1.43\times 10^{-5}\mr{m}^{-1}$ gives
}
\begin{equation}
\Ro = 0.4, \ \ k_*\Delta=5, \ \ l_*\Delta=5, \ \ J=5 \ \ \textrm{and} \ \  \nu_*=1,
\label{param2}
\end{equation}
so $r =k_*\Delta \Ro =2$. {Since the parameters $\Ro^{-1}$, $k_*\Delta$ and $J$ are only moderately large, this choice provides a strict test on the applicability of the asymptotic results.}
The results for the form of the wavepacket are shown in Fig.\ \ref{FigWave2}. The accuracy of the asymptotic approximations has degraded considerable compared with that in Fig.\ \ref{FigWave2}, unsurprisingly, perhaps, given that the error scales like {$(k_* \Delta)^{-1/2}\approx0.45$}. Nonetheless, there remains a reasonable qualitative match between asymptotic and numerical results which suggests our approximations remain useful.

\section{Wave--mean-flow interaction}\label{Sec_wm}

Section \ref{wave_solu} shows that the wave amplitude changes suddenly across the  inertial layer. In this section, we exploit our asymptotic expression for the wavepacket structure to derive the mean force exerted in this layer as a result of this change,  and to calculate the mean-flow response.
The important quantity for this force is the EP flux, which has long been recognized as the relevant diagnostic \citep{Elia1961,Andr1976,Boyd1976,Edmo1980}.

Two types of mean-flow response need to be distinguished: the far-field response, and the local response.
The far-field response is the net change in the mean flow that persists far  downstream of the mountain; 
it is a consequence of a change of the horizontally integrated EP flux due to absorption.
In contrast, the local response is the mean-flow change caused by local EP flux changes without far-field impact because they integrate to zero horizontally. As discussed in  \S\ref{wave_solu}, the net EP flux change is concentrated in the inertial layer where the Regime II$_\mathrm{B}$ approximation applies; below this, the waves are {localized} horizontally in space and leave no net mean-flow response, in agreement with non-acceleration results.
%: the mean flow generation only exists within the range of wavepacket but not outside.
%
%
%
% as is the case for the (cancelling) changes at the leading and trailing edges of the wavepacket. \commentJV{Do you agree with the sentence above?}. We focus on the far-field response. 
%
%
%As discussed in  \S\ref{wave_solu}, the net EP flux change is concentrated in Regime II$_\mathrm{B}$ denoting the inertial layer; for the rest of the region, waves are transient in space that leave no net mean-flow response: the mean flow generation only exists within the range of wavepacket but not outside.
%In accord with the non-acceleration theorem, if zonal (streamwise) integration is taken, this local mean-flow generation does not appear.
In the remainder of this section, we derive the equation governing mean-flow response (\S\,\ref{SecGE}), compute the EP flux divergence that appears as the sole forcing term in this equation (\S\,\ref{SecEP}), and compute the mean-flow change asymptotically, taking advantage of the the thinness of the inertial layer (\S\,\ref{SecPV})

\subsection{Governing equation} \label{SecGE}

Taking advantage of the small Rossby number, the mean flow is calculated using quasi-geostrophic theory.
By taking the horizontal curl of the horizontal momentum equation (\ref{Primomh}), applying $\partial_zf/N^2$ to (\ref{Pript}) and using the incompressibility (\ref{Priincomp}), we obtain
\begin{eqnarray}
\partial_t \br{v_x-u_y + \partial_z\br{\frac{f}{N^2} b}} + \underbrace{(\partial_{xx}-\partial_{yy})(uv) + \partial_{xy}(v^2-u^2)}\limits_{(\mathrm{a})} + \underbrace{\partial_{z}\br{\frac{f}{N^2}\partial_z(wb)}}\limits_{(\mathrm{b})} \nonumber\\
+ \underbrace{\partial_{xz}\br{ wv+\frac{f}{N^2}ub } - \partial_{yz}\br{ wu - \frac{f}{N^2}vb }}\limits_{(\mathrm{c})} = 0. \label{MeanEq}
\end{eqnarray}
The waves and mean flow are separated by the small-scale average defined as
\begin{equation}
\av{~\cdot~} = \frac{1}{D^2}\int_x^{x+D}\int_y^{y+D} \cdot~ \dd \bx', \label{average}
\end{equation}
where $ k_*^{-1} \ll D \ll \Delta$.
Because of their small spatial scale, the waves have zero average.
Applying this  average to (\ref{MeanEq}), and using the smallness of the wave amplitude and Rossby number to retain the leading order terms for both wave and mean flow, we obtain
\begin{equation}
(\partial_t + \Lambda z \partial_x )\br{\nabla^\perp\cdot\bU + \partial_z\br{\frac{f}{N^2}\av{b_2}}} + \nabla^\perp\cdot\partial_z \bF = 0, \label{wm_dim}
\end{equation}
where $\nabla^\perp = (-\partial_y,\,\partial_x)$ denotes the horizontal curl, and
\begin{equation}
\bF = \br{ \av{u_1w_1}-\frac{f}{N^2}\av{v_1b_1},\,\av{v_1w_1}+\frac{f}{N^2}\av{u_1b_1}} \label{defEP}
\end{equation}
is the vertical part of the Eliassen--Palm (EP) flux \cite{Elia1961}. In (\ref{wm_dim}), 
the  mean flow is in geostrophic balance, with 
\begin{equation}
\langle \bu_2 \rangle = (\bU,0) = (U,V,0) = (-\partial_y \Psi,\, \partial_x \Psi,0) \quad \textrm{and} \quad  \av{b_2} = f\partial_z\Psi,
\end{equation} 
where the streamfunction $\Psi$ remains to be determined.

If there is no wave effect, $\bF=0$, (\ref{wm_dim}) reduces to the quasi-geostrophic potential-vorticity (QGPV) equation.
Note that term (c) in (\ref{MeanEq}) is the only quadratic wave term to contribute to (\ref{wm_dim}): 
term (a) is $O(\delta_*/H)$ smaller than term (c) because of the vertically thinness of inertial layer;
term (b) has zero average because $w_1$ and $b_1$ are out of phase in the limit of large $J$ (see (\ref{hatb})).

The  boundary condition associated with (\ref{wm_dim}) is obtained by taking the average of (\ref{bc}),
retaining terms  up to second order in the wave amplitude to find 
\begin{align}
\av{w_2} &= \av{ \mb{u}_1\cdot \nabla h_\mathrm{t} - h_\mathrm{t}\partial_{z} w_1 - \Lambda h_\mathrm{t}\partial_x h_\mathrm{t}} 
= \av{ \mb{u}_1\cdot \nabla h_\mathrm{t} + h_\mathrm{t} \nabla \cdot \mb{u}_1 - \Lambda h_\mathrm{t}\partial_x h_\mathrm{t}} \nonumber \\
&= \av{\nabla\cdot(\mb{u}_1 h_\mathrm{t}) - \Lambda \partial_x \br{{h_\mathrm{t}^2}/{2}} }= 0, \label{MeanWbc}
\end{align}
where $\mb{u}_1 = (u_1,\,v_1)$, and the incompressibility condition (\ref{wave_incom}) is used.
%Here $h_\mathrm{t}\partial_{z} w_1$ is the Stokes drift and physically speaking condition 

One significant feature of the EP flux is its conservation when the background flow possesses certain symmetries, leading to the non-acceleration theorem \citep{Char1961}: waves do not force the mean flow unless there exists a singularity or some dissipation. In our setup, the background shear flow has $x$- and $y$-symmetry, and  EP flux conservation is easily demonstrated for a plane wave with wavevector $\bk=(k,l)$: applying the polarization relation (\ref{PolRel}) to (\ref{defEP}), we obtain the plane-wave expression of the EP flux,
\begin{equation}
\bF = (F,\nu F), \quad \textrm{with} \quad F = \frac{-\Lambda}{f}\frac{1}{1+\nu^2} \Re\left( \ii \frac{1-\zeta^2}{\zeta^2} \hat{w}_{\zeta}(\bk) \hat{w}^*(\bk) - \nu \frac{\hat{w}(\bk) \hat{w}^*(\bk)}{\zeta^2}  \right). \label{EPk}
\end{equation}
Its conservation is deduced from (\ref{WaveEq}) by multiplication of $\hat{w}$ by its complex conjugate $\hat{w}^*$ and  subtraction of the conjugate of the resulting equation to find
\begin{eqnarray}
\partial_z F = 0. \label{singleEP}
\end{eqnarray}
%Here, only $x$-component EP flux in wavenumber space is considered, and $y$-component EP flux equals $\nu$ times the $x$-component EP flux.
%As is pointed out in Appendix \ref{Sec_sSA}, because of the scale separation the horizontal integrated EP flux to the leading order equals to the integration of EP fluxes of wavenumber such that $\overline{F} =  \int_{-\infty}^{\infty}\int_{-\infty}^{\infty}\breve{F}(\bk)\dd \bk$.
This conservation does not hold across the inertial level singularities $\zeta=\pm 1$. Across the lower one, $\zeta=1$, the plane wave EP flux attenuates to an exponentially small value, leading to the wave forcing of the mean flow in the inertial layer (Regime II$_\mathrm{B}$). We now compute the EP flux in this layer. 

%As to the specific scales, the vertical and horizontal EP flux divergence have the typical orders of $\partial_z\av{w_1u_1}$ and $\partial_x\av{u_1v_1}$, respectively, and orders of the partial derivatives should be estimated by the characteristic scales of the mean effect.
%If there is no singularity, these two terms are of the same order as
%\begin{equation}
%(k_*\Delta)^2\br{\frac{h}{H}}^2\frac{U_\mathrm{b}}{\Delta}.
%\end{equation} 
%When the inertial level exist, the correspondent inertial layer has the characteristic scale $\delta_*$ (see (\ref{ilscale})), thus the vertical EP flux is of $O(H/\delta_*)$ larger than the horizontal components.
%Thus, in leading equation (\ref{wm_dim}) only the vertical components of EP flux remain due to the large vertical derivative brought about by the inertial-level singularity.

\subsection{Eliassen--Palm flux} \label{SecEP}

The derivation of the EP flux is greatly simplified by observing that, in the saddle-point approximations valid in Regimes I and II, the relations of the various wave fields $u_1$, $v_1$, etc.\ associated with the wavepacket to $w_1$ mirror the polarisation relations (\ref{PolRel}). This is because the rapid dependence in $x$ of the wave solution corresponds to a plane wave with (possibly complex) wavenumber $\bk_\mathrm{s}$ given by a saddle-point value of $\bk$ (see Appendix \ref{appendix1}). 
%
%
%In (\ref{wm_dim}) the wave and the mean flow interact through EP flux defined by (\ref{defEP}). 
%As a wave quantity, it is calculated using approximate wave expressions obtained in \S \ref{wave_solu}.
%Since $x$ and $y$ component EP flux differs by a factor $\nu=\nu_* + \mathrm{h.o.t.}$, which to the leading order is a same constant for all wavenumbers, we concentrate on calculating the $x$-component EP flux.
%
%It can be seen from the $w_1$ approximations that by applying the steepest descent method to a wavepacket its expression is controlled by a single frequency (complex) -- the saddle point $K_\mathrm{s}$.
%In Regime I $K_\mathrm{s}=k_*$ is a special case where ray tracing validates. 
Using the notation $\zeta_\mathrm{s}=-k_\mathrm{s}\Lambda z/f$, 
%and the functions in spectrum space by their inverse Fourier transforms, 
we obtain
\begin{subequations}\label{PolRel2}
	\begin{align}
	u_1 &=  \frac{\ii(\zeta_{\mathrm{s}} - \ii \nu_*)}{K_\mathrm{s}\zeta_{\mathrm{s}}(1+\nu_*^2)} w_{1z} - \frac{\ii \Lambda\nu_*^2}{f\zeta_{\mathrm{s}}(1+\nu_*^2)} w_1 ,\\
	v_1 &=  -\frac{1 - \ii \nu_*\zeta_{\mathrm{s}}}{K_\mathrm{s}\zeta_{\mathrm{s}}(1+\nu_*^2)} w_{1z} + \frac{\ii\Lambda\nu_*}{f\zeta_{\mathrm{s}}(1+\nu_*^2)} w_1 ,\\
	b_1 &= -\frac{\ii \Lambda(1 - \ii \nu_*\zeta_{\mathrm{s}})}{K_\mathrm{s}\zeta_{\mathrm{s}}^2(1+\nu_*^2)} w_{1z} + \ii\frac{\Lambda^2}{f}\br{ \frac{\ii \nu_*}{\zeta_{\mathrm{s}}^2(1+\nu_*^2)}  + \frac{J^2}{\zeta_{\mathrm{s}}}}  w_1,
	\end{align}
\end{subequations}
where we also have used that $\nu_\mathrm{s} = l_\mathrm{s}/k_\mathrm{s} \sim l_*/k_* = \nu_*$ to leading order. 
%where 
%\begin{equation}
%w_{1z} = \mathcal{F}
%\end{equation}
Correspondingly, a derivation that parallels that of (\ref{EPk}) gives the $x$-component of the wavepacket EP flux as
\begin{equation}
F = \frac{1}{1+\nu_*^2}  \Re\left( \frac{\ii}{k_\mathrm{s}} \frac{1-\zeta_{\mathrm{s}}^2}{\zeta_{\mathrm{s}}^2} w_{1z} w_1^* - \nu_* \frac{w_1 w_1^*}{\zeta_{\mathrm{s}}^2}  \right). \label{FF}
\end{equation}

This can be simplified further. Focussing on Regime II$_\mathrm{B}$, we observe that the $O(\epsilon)$ vertical scale implies that the first term in the brackets in (\ref{FF}) dominates the second, that $k_\mathrm{s}$ can be approximated by $k_*$, and that $\zeta_\mathrm{s}$ can be approximated by $1$ except in the factor $1-\zeta_\mathrm{s}$ of $1-\zeta_{\mathrm{s}}^2$. This leads to the simple expression
\begin{equation}
F = \frac{2(1-\zeta_{\mathrm{s}})}{k_*(1+\nu_*^2)} \Re\left( \ii  w_{1z} w_1^*  \right). \label{IIBEP} 
\end{equation}
Now, using that $\partial_z= -k^* \Delta \Lambda \partial_Z /f$, we obtain from the form of $w_1$ in 
(\ref{fullw1IIB}) that, to leading order,
\begin{equation}
w_{1z} \sim \frac{\ii (k_* \delta)^{3/2} k_* \Lambda}{f} w_1.
\end{equation}
Introducing this result into (\ref{IIBEP}) and using that 
\begin{equation}
\zeta_\mathrm{s} = 1+ \frac{1}{k_*\Delta}(K_\mathrm{s}+Z),
\end{equation}
together with the explicit form (\ref{SP}) of $K_\mathrm{s}$, we find that
\begin{equation}
F = \frac{J^2 \Lambda}{(k_* \Delta)^{3/2}f X} |w_1|^2.
\end{equation}
The explicit form of $w_1$ in (\ref{wB}) can finally be used to obtain the explicit expression
%
%
%As we have already discussed that the net change of EP flux by wave absorption is only important in the inertial layer represented by the Regime II$_\mathrm{B}$, so the relation (\ref{PolRel2}) can be simplified to
%\begin{subequations}\label{PolRel3}
%\begin{align}
%u_1 &=  \frac{\ii(\zeta_{\mathrm{s}} - \ii \nu_*)}{K_\mathrm{s}\zeta_{\mathrm{s}}(1+\nu_*^2)} w_{1z},\\
%v_1 &=  -\frac{1 - \ii \nu_*\zeta_{\mathrm{s}}}{K_\mathrm{s}\zeta_{\mathrm{s}}(1+\nu_*^2)} w_{1z},\\
%b_1 &=  \ii\frac{\Lambda^2}{f} \frac{J^2}{\zeta_{\mathrm{s}}}  w_1,
%\end{align}
%\end{subequations}
%leading to the approximation of EP flux
%
%Here follow the calculation of $w_1$ in Appendix \ref{RegimeIIB} using stationary phase method we obtain
%\begin{equation}
%w_{1z} \doteq \ii \Lambda h J^{1/2}k_*(k_*\Delta)\br{r^2-1}^{1/4} \br{1+\nu_*^2}^{1/4}  \ex^{-\frac{1}{2\Delta^2}\br{y - \frac{J \nu_*}{k_*\sqrt{1+\nu_*^2}}D_\mathrm{II} }^2} \ex^{ -\frac{1}{2}\br{\frac{J^2\br{1+\nu_*^2}}{2(k_*\Delta)^{2}X^2}-Z}^2 }.
%\end{equation}
%Combining with 
%
%where $K_\mathrm{s}$ is expressed in (\ref{SP}),
%the $x$-component EP flux (\ref{IIBEP}) become
\begin{equation}
F =   \frac{E}{X^3} \, \ex^{ -\left(a^2/X^2-Z\right)^2} g^2(y), \label{EP}
\end{equation}
where
\begin{equation}
E = \frac{J^{3}\Lambda h^2f \br{r^2-1}^{1/2} \br{1+\nu_*^2}^{1/2}}{(k_*\Delta)^{5/2} }
\end{equation}
is a constant controlling the amplitude of $F$.
This expression only holds for $X>0$; for $X<0$, the EP flux is exponentially small.
We emphasise the remarkably simple form of (\ref{EP}): notwithstanding the many parameters involved, the formula is well suited for practical use in parameterisations.
%so taking $\zeta$ derivative in (\ref{w}) and approximating in Regime II$_\mathrm{B}$ we obtain to the leading order that
%\begin{equation}
%\ii \frac{1-\zeta^2}{\zeta^2} \hat{w}_{\zeta} = -\epsilon^{-1/2}(k_*\Delta)^{1/2}\sqrt{1+\nu_*}\sqrt{2(K+Z)}. \hat{w} \label{ue0}
%\end{equation}
%Since (\ref{ue0}) is of $O((k_*\Delta)^{-1/2})$ larger than $\nu \hat{w}/\zeta^2$, to the leading order (\ref{EPk}) become
%\begin{eqnarray}
%F(\bk) = \frac{-\Lambda}{f}\frac{1}{1+\nu^2} \Re\left\{ \ii \frac{1-\zeta^2}{\zeta^2} \hat{w}_{\zeta}(\bk) \hat{w}(\bk)^* \right\}.
%\end{eqnarray}
%
%As is pointed out in Appendix \ref{Sec_sSA}, we can calculate the small scale averaged EP flux from the single wavenumber EP flux, so
%\begin{equation}
%\F = \frac{1}{2}|u_{\mathrm{e}}||w_1|, \label{EP_def}
%\end{equation}
%where
%\begin{equation}
%u_{\mathrm{e}} = \frac{-\Lambda}{f}\frac{1}{1+\nu_*^2}\int_{-fz^{-1}\Lambda^{-1}}^{\infty} \int_{-\infty}^{\infty} \ii \frac{1-\zeta^2}{\zeta^2} \hat{w}_{\zeta} \dd k\dd l,
%\end{equation}
%Here, we have considered that $u_{\mathrm{e}}$ and $w_1$ are ``in phase'' indicated by (\ref{ue0}).
%
%In regime II$_\mathrm{B}$, similar as $w_1$, by applying the stationary phase method $u_{\mathrm{e}}$ is approximated as 
%\begin{equation}
%u_{\mathrm{e}} = \frac{\ii \Lambda h J^{5/2}\br{r^2-1}^{1/4} \br{1+\nu_*^2}^{1/4}}{(k_*\Delta)^2 } \ex^{-\frac{1}{2\Delta^2}\br{y - \frac{J \nu_*}{k_*\sqrt{1+\nu_*^2}}D_\mathrm{II} }^2}
%\frac{1}{X^2} \ex^{ -\frac{1}{2}\br{\frac{J^2\br{1+\nu_*^2}}{2(k_*\Delta)^{2}X^2}-Z}^2 }.
%\end{equation}
%So from (\ref{EP_def}) we obtain 

\begin{figure}
	\centering
	\includegraphics[width=0.9\textwidth]{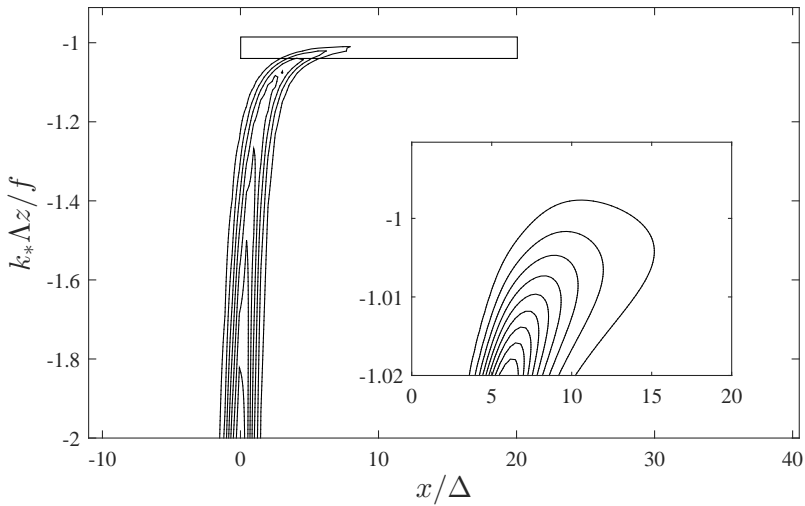}
	\caption{Contours of the EP flux in (\ref{EP}) normalized by the largest value at the bottom $(3/4)^{3/4}\ex^{-3/4}E$ for the parameters in (\ref{param1}); 7 equispaced contours in the range $[0.125,0.875]$ are shown in the main figure, and in the inset, 9 equispaced contours are in the range $[0.03,0.27]$.
		The contours can be compared  with those of the vertical velocity $|w_1|$ shown in Fig.\ \ref{FigContour}: although the two fields obey the same scaling, the maximum of the EP flux is closer to the mountain. The inset zooms on the rectangular box indicated in the main panel.} %with equispace; in the inset, 7 contours ranges from $0.03075$ to $0.1845$ with equispace.}
	\label{FIG_epcontour}
\end{figure}

\begin{figure}
	\centering
	\includegraphics[width=0.6\textwidth]{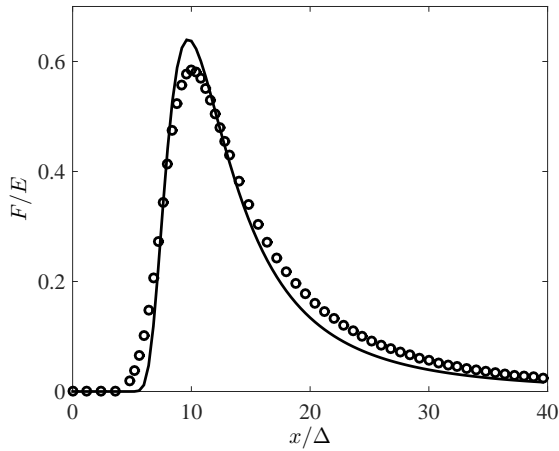}
	\caption{EP flux at the dominant inertial level for the parameters in (\ref{param1}): numerical results (circles) are compared with the prediction of the Regime II$_\mathrm{B}$ approximation (solid line).}
	\label{fig_ep}
\end{figure}

We illustrate the form of the EP flux in Fig.\ \ref{FIG_epcontour} by showing its contours in the $(x,z)$ cross-section where it is maximised, using the parameters in (\ref{param1}).
%$\Ro = 0.02$, $k_*\Delta=100$, $l_*\Delta=100$, $J=100$ and $\nu_*=1$ . 
The EP flux is computed from the WKB linear solution (\ref{WOri0}) and polarization relation (\ref{PolRel}).
The validity of the asymptotic approximation is confirmed by Fig.\ \ref{fig_ep} which compares the EP flux obtained numerically with the asymptotic approximation at  the dominant inertial level.

By integrating (\ref{EP}) over $x$ and $y$, we obtain the horizontally integrated EP flux
\begin{eqnarray}
\overline{F} = \frac{\pi J \Delta^2 \Lambda h^2f \br{r^2-1}^{1/2} }{2\br{1+\nu_*^2}^{1/2}} (1+\mathrm{erf}Z), \label{Epvar}
\end{eqnarray}
where the decay of the error function to $0$ as $Z \to -\infty$ (above the inertial layer) clearly captures wave absorption. As $Z \to \infty$, i.e., well below the dominant inertial level, $\overline{F}$ tends to the constant value
\begin{equation}
\overline{F}_{\mathrm{tot}} = \frac{\pi J \Delta^2 \Lambda h^2f \br{r^2-1}^{1/2} }{\br{1+\nu_*^2}^{1/2}}. \label{EPt1}
\end{equation}
This matches the horizontally integrated EP flux at the ground, since the flux is conserved below the inertial layer. We check this in Appendix \ref{app:bottom}. We note that $\overline{F}_{\mathrm{tot}}$ is finite in the  limit $f \to 0$, where it is given by $\overline{F}_{\mathrm{tot}} = \pi N \Lambda \Delta^2 h^2 k_* H$. {The integrated EP flux $\overline{F}$ becomes discontinuous in this limit, jumping from $\overline{F}_{\mathrm{tot}}$ to $0$ across the critical level.}

 % \commentJHX{with $\pi\Delta^2$ comes from the envelope of the mountain.}, recovering the value found in the non-rotating case. 
%\commentJV{Could you check the non-rotating case? I've done it starting with p152 of Oliver's book, but I'm not too sure how the factors $\pi$ and $\Delta$ work out.}
%\commentJHX{$\pi\Delta^2$ has a good geometry meaning that it represents the area of mountain envelope $\ex^{-(x^2+y^2)/(2\Delta^2)}$. I feel that checking the total EP flux expression is not physically interesting, since it is fixed by the bottom boundary regardless how waves break or absorbed. But it is good to know that we have right coefficients.}
%\commentJHX{I could not find expression for non-rotating critical result, and I think it is reasonable since all wavenumbers have the same critical level and it is hard to say the vertical structure based on a linear theory. For single wavenumber expression I think it dates back to 1969 Bretherton's paper, and the result is trivially expressed by the boundary $\av{wu}$ because of EP-conservation.}
From (\ref{Epvar}), we can also estimate the wave drag per unit area, defined as the vertical derivative of horizontally integrated EP flux divided by $\Delta^2$:
\begin{equation}
\overline{f(Z)} = \frac{\partial_z\overline{F(Z)}}{\Delta^2} = 
-\frac{\sqrt{\pi}}{2}\Lambda N \Delta h^2 k_*^2 
\frac{k_*}{|\boldsymbol{\tk}|}(r^2-1)^{1/2}\ex^{-Z^2} , \label{EPfor_dim}
\end{equation} 
where $|\boldsymbol{\tk}| = \sqrt{\tk^2+\tl^2}$ is the amplitude of dominant wavenumber.

\subsection{Mean-flow response} \label{SecPV}

We now consider the mean-flow response to the wave drag associated with the $z$-dependent EP flux (\ref{EP}). We compute the steady flow response by solving the QGPV equation (\ref{wm_dim}) asymptotically, taking advantage of the thinness of the inertial layer where the wave drag acts to apply matched asymptotics. Thus the domain $z>-H$ is separated into an inner region around the inertial level $z \approx z_*$, specifically $Z=-k_*^2 \Delta \Lambda (z-z_*)/f=O(1)$, and an outer region where the wave drag is absent. Simplifications arise because the spatial scale in the $x$-direction is longer than in the $y$-direction; this is made explicit using the variable $X$ defined in (\ref{X}), with $X=O(1)$ implying that $x=\Delta \times O((k_* \Delta)^{1/2})$

We denote the streamfunction associated with the wave-induced mean flow in the inner and  outer regions by $\Psi(X,y,Z)$ and $\psi(X,y,z)$, respectively. In the inner region, considering the scalings of Regime II$_\mathrm{B}$, the steady ($\partial_t = 0$) QGPV equation (\ref{wm_dim}) becomes 
\begin{align}
& \epsilon^{1/2}k_*(k_*\Delta)^{-3/2}(U_*\partial_X + \nabla^{\perp} \Psi \cdot \nabla) \left(\epsilon k_*^2(k_*\Delta)^{-3} \partial_{XX} \Psi + \partial_{yy} \Psi + \frac{k_*^4\Delta^2}{\epsilon^2J^2}\partial_{ZZ}\Psi\right) \nonumber \\
& \qquad \qquad \qquad \qquad  = \frac{k_*^2\Lambda\Delta}{\epsilon^2 f}\partial_{yZ}F, \label{dpv_in}
\end{align}
where $\nabla$ and $\nabla^\perp$ are gradients with respect to the scaled variables $(X,y)$, $U_*=-\Lambda \tz = f/k_*$ is the background velocity at the dominant inertial level, and we have included the bookkeeping parameter $\epsilon$. On the right-hand side we have neglected the $x$ derivative of $\partial_z F$ against the $y$-derivative, owing to the asymptotically larger scales in $x$.  

For sufficiently small mountain height, Eq.\ (\ref{dpv_in}) can be linearised; we make explicit below the condition for this approximation to hold. Retaining only the leading-order terms in  (\ref{dpv_in}) reduces this to
\begin{equation}
\frac{k_*^3(k_*\Delta)^{1/2}}{\epsilon^{3/2}J^2} U_* \partial_{XZZ}\Psi  = \frac{k_*^2\Lambda\Delta}{\epsilon^2 f}\partial_{yZ}F. \label{eq_in}
\end{equation}

Since $F$ is exponentially small for $X<0$, we can integrate (\ref{eq_in}) for $X>0$ to obtain
\begin{equation}
\partial_{ZZ} \Psi = Q, \label{Q_eq}
\end{equation}
where 
\begin{equation}
Q = Q(X,y,Z) = \frac{J^2\Lambda(k_*\Delta)^{1/2}}{\epsilon^{1/2}U_* k_*^2 f} \int_{0}^{X} \partial_{yZ}F(X',y,Z) \, \dd X' \label{PVchange}
\end{equation}
can be interpreted as a scaled wave-induced PV. This can be computed explicitly using (\ref{EP}) to find
\begin{equation}
Q = \frac{J^3 \Lambda^2 \Delta h^2 (r^2 -1)^{1/2}}{f (k_* \Delta) (1+\nu_*^2)^{1/2}} \, \ex^{-(a^2/X^2-Z)^2} \partial_y g^2,
\end{equation}
with $g(y)$ defined in (\ref{g(y)}).

Eq.\ (\ref{Q_eq})  is readily integrated, leading to
\begin{equation}
{\Psi}(X,y,Z) = \int_{0}^{Z}\int_{0}^{Z'}Q(x,Y,Z'')\,  \dd Z' \dd Z'' + C_1(X,y) Z + \epsilon^{-1} C_2(X,y),
\end{equation}
where $C_1$ and $C_2$ are integration `constants' that are determined by matching the outer solution {\citep[cf.][]{plou-v}}. We have anticipated that the $Z$-independent term is an order $\epsilon^{-1}$ larger than the other terms. Matching requires the asymptotic behaviour of $\Psi$ as $Z \to \pm \infty$, found to be
\begin{align}
{\Psi}(X,y,Z) \sim& Z \left(\int_{0}^{\pm \infty}Q(X,y,Z')\, \dd Z' + C_1(X,y) \right) + \epsilon^{-1} C_2 (X,y) \nonumber
\\  & - \int_0^{\pm\infty} Z' Q(X,y,Z') \, \dd Z'  \label{solu_in1}
\end{align}
as $Z \to \pm \infty$. %\commentJV{I introduced $\epsilon$ is this, otherwise it was not clear that $C_2^\pm$ are just small corrections to $C_2$.}

In the outer region, the QGPV equation (\ref{wm_dim}) is
\begin{equation}
\Lambda z \partial_X  \left(\partial_{yy}\psi + \epsilon k_*^2(k_*\Delta)^{-3} \partial_{XX}\psi + \frac{f^2}{N^2}\partial_{zz}\psi\right) = 0,
\end{equation}
which, to the leading order, reduces to
\begin{equation}
\Lambda z \partial_X (\partial_{yy}\psi + \frac{f^2}{N^2}\partial_{zz}\psi) = 0. \label{eq_out}
\end{equation}
Integrating in $X$, we find that
\begin{equation}
\partial_{yy}\psi + \frac{f^2}{N^2}\partial_{zz}\psi = 0.
\end{equation} 
This is best solved using a Fourier transform in the $y$ direction. Denoting this transform by a hat, we have
\begin{equation}
\hat{\psi} =
\begin{cases}
\epsilon^{-1} \hat C_3(X,l) \ex^{-N|l|(z-z_*)/f} & \textrm{for} \ \ z>z_*\\
\epsilon^{-1} \hat  C_4(X,l) \ex^{-N|l|(z-z_*)/f} + \epsilon^{-1} \hat C_5(X,l) \ex^{N|l|(z-z_*)/f} & \textrm{for} \ \  z<z_* \label{solu_out} 
\end{cases},
\end{equation}
applying a vanishing boundary condition as $z\to \infty$.
%\footnote{We can also apply other boundary conditions above the singularities, but extra attention need to be paid if the upper boundary condition supports downwards propagating waves. In addition we need to consider correspondent wave solutions.}
Combining the condition (\ref{MeanWbc}) of zero mean vertical velocity with the buoyancy equation $U\partial_X \psi_z = 0$ stemming from (\ref{Pript}), we find the condition  $\psi_z = 0$ at the lower boundary $z=-H$. 
This implies
\begin{equation}
\hat C_4 \ex^{N|l|(H+z_*)/f} =  \hat C_5 \ex^{-N|l|(H+z_*)/f}. \label{PVbc}
\end{equation}

We now match (\ref{solu_out}) to (\ref{solu_in1}) to determine $\hat C_3$, $\hat C_4$ and $\hat C_5$, and hence the outer solution, completely. Substituting $z-z_* = - \epsilon  f Z /(k_*^2 \Delta \Lambda)$ into (\ref{solu_out}), expanding in powers of $\epsilon$ and matching with (\ref{solu_in1}), we find that
\begin{equation}
\hat C_4+ \hat C_5= \hat C_3 \, (= \hat{C}_2), \label{conti_cond}
\end{equation}
and
\begin{align}
\hat C_4  - \hat C_5 - \hat C_3 &= \frac{k_*^2\Delta}{J|l|}\int_{-\infty}^{\infty}\hat{Q}(X,l,Z')\,\dd Z' \nonumber\\
&= \frac{\sqrt{\pi}J^2 (k_* \Delta) \Lambda^2 h^2  (r^2 -1)^{1/2}}{f (1+\nu_*^2)^{1/2}} \frac{\widehat{\partial_y g^2}}{|l|}. \label{matching}
\end{align}
Thus, (\ref{PVbc}), (\ref{conti_cond}) and (\ref{matching}) provides three equations for $\hat C_3$, $\hat C_4$ and $\hat C_5$ and hence determine the mean flow. The solution to these equations is straightforward but leads to lenghty expressions which we relegate to Appendix \ref{SecEg}.

Remarkably, the right-hand side of (\ref{matching}) and hence $\hat C_3$, $\hat C_4$ and $\hat C_5$ do not depend on $X$. As a result, $\psi$, $\Psi$ and thus the entire mean-flow response does not change downstream of the mountain. Our solution suggests that there is a jump in this response, from a zero value for $X<0$ to the $X$-independent value for $X>0$. This is an artefact of the asymptotic approximation: the transition to a non-zero mean flow is in fact smooth. Its detailed form could be obtained using the  approximation of the wave fields in Regime II. Here we only note that the scaling of Regime II indicates that the transition region has a characteristic length $x/\Delta=O((k_*\Delta)^{1/3})$, asymptotically smaller than the $O((k_* \Delta)^{1/2})$ scale that is resolved by the Regime IIB approximation used in our computation of the mean-flow response, hence the apparent discontinuity.

Eq.\ (\ref{matching}) provides an estimate for the order of magnitude of the mean-flow response. Recalling that 
the change in mean velocity is $- \partial_y \psi$ with the $y$-scale $\Delta$, and noting that the maximum value of $\psi$ is $O(\hat C_3)$, we estimate the wave-induced mean velocity as
\begin{equation}
U_\mathrm{w} = O\left(J^2 r \left(\frac{h}{H}\right)^2 U_\mathrm{b} \right),
\label{U_w}
\end{equation}
{assuming that $(r^2-1)^{1/2} = O(1)$.}
Since $(k_* \Delta) \Ro = O(1)$, this indicates that $U_\mathrm{w} \ll U_\mathrm{b}$, as required for the linearisation of the QGPV equation, provided that $J(h/H) \ll 1$.

\begin{figure}
	\hspace{-1.4cm}
	\includegraphics[width=1.2\textwidth]{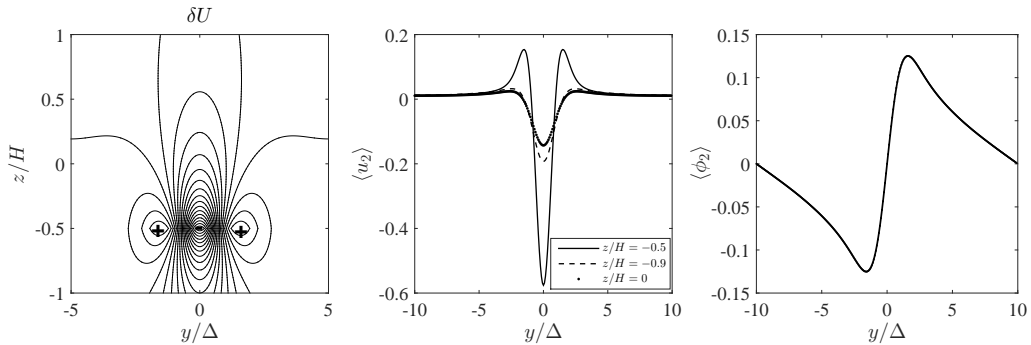}
	\caption{Mean-flow response. Left: contours of the wave-induced mean flow $\langle u_2 \rangle$, with the sign is indicated by $\pm$; 20 equispaced  contours in the range $[-0.52, 0.13]$ are shown. Middle: $\langle u_2 \rangle$ at $z/H=-0.9,\,-0.5,\,0$, with  $z=-0.5$ corresponding to the dominant inertial level. The mean flow is normalized by $ \sqrt{3\pi} k_* J^2 \Lambda^2 h^2/f$. {Right: mean pressure $\langle \phi_2 \rangle$ at ground level ($z=-H$), normalized by $ \sqrt{3\pi} k_*\Delta J^2 \Lambda^2 h^2$.} }
	\label{mean}
\end{figure}

To illustrate our results, we have calculated the mean-flow response for $J=k_*\Delta$, $l_*=0$ and $r=2$. %\commentJV{I don't understand why, if $J=k_*\Delta=1$, you kept $J$ and $k_* \Delta$ in the expressions below for $C_3$ etc. Is the only approximation that $l_*=0$?}
The linear system (\ref{PVbc}), (\ref{conti_cond}) and (\ref{matching}) is readily solved for $\hat C_3$, $\hat C_4$ and $\hat C_5$, leading to $\hat \psi$ and, after Fourier inversion, to the mean flow $\langle u_2 \rangle = - \partial_y \psi$ {and mean pressure $\langle \phi_2 \rangle$}. This is displayed in Fig.\ \ref{mean}.   
Observe that the mean-flow response to the wave drag localised in the thin inertial layer is distributed through the entire depth of the fluid, and that the total mean-flow change $\int \langle u_2 \rangle \, \dd y$ vanishes at each altitude since the streamfunction $\psi$ vanishes as $y \to \pm \infty$. 
%\commentJHX{At the bottom topography, mean flow accompanies an antisymmetric surface pressure which takes its extremes at the positions with zero mean velocity due to geostrophy balance.}
%For any given altitude the extreme values of the velocity change locate at the middle ($y=0$) and at the position close to the boundary of the wave-force region (see the right panel in Figure \ref{mean}). 
%Since the extreme values have opposite signs, horizontal mean shear is generated.

{In the limit of $f\to 0$, i.e. $r \to \infty$, (\ref{matching}) indicates that $U_\mathrm{w} \to \infty$. This is because the quasi-geostrophic approximation breaks down: the wave-forcing associated with the jump in the EP flux cannot be balanced by the Coriolis force but instead leads to an acceleration of the mean flow. A more meaningful limit treats $f$ as finite for the mean-flow response but sets it to $0$ for the evaluation of the EP flux. The fact, noted in \S\ref{sec:saddles}, that the EP flux undergoes a discontinuous jump in this case has little impact on the mean-flow response since, with the QGPV equation (\ref{wm_dim}) remaining valid, the mean flow away from the inertial layer  depends only on the magnitude of the jump and not on the details of its structure. 
%\commentJHX{The QGPV  validates to predict the magnitude of wave induced mean flow.}
Note however that the right-hand side of (\ref{U_w}) needs to multiplied by an extra factor $r$ to account for the fact that $(r^2-1)^{1/2} \sim r \gg 1$.}

%\commentJV{I propose to remove all the rest, up to the Conclusion section. We however need to say something about the limit $f \to 0$.}
%
%\commentJV{I left the par below for your to verify which of my estimate or yours is the correct one. Maybe they both are.}
%and their orders can be estimated with the help of the order of EP flux (\ref{EP}) that 
%\begin{equation}
%C_i = O\br{\frac{J^4}{(k_*\Delta)^2}\br{\frac{h}{H}}^2 U_\mathrm{b}\Delta^2 } \quad (i=3,4,5). \label{order_C}
%\end{equation}
%So to validate the linearized PV equations (\ref{eq_in}) and (\ref{eq_out}), $C_i \ll 1$ is required, and it is ensured in our scaling that $J=O( k_*\Delta)$ and $J(h/H) \ll 1$.
%Eq. (\ref{order_C}) also shows that the mean-flow generation is a second order effect, which is in consistence with the energy conservation argument.
%
%\commentJV{I propose to get rid of the example section below and just keep the figures: the explicit form of the $C_i$ is not particularly useful, and the prefactors are now worked out in the previous section. What do you think?} 

\section{Conclusion} \label{conclusion}

In this paper, we study the propagation of a mountain IGW wavepacket in a rotating shear flow and the mean flow  generated as a result of wave absorption at inertial-level singularities. 
The broad wavenumber spectrum of the wavepacket and the dependence of the inertial-level altitude on wavenumber lead to a smearing-out of the singularities over a finite-thickness inertial layer where the mean-flow forcing concentrates. Thus, in contrast with the situation when rotation is neglected, dissipative processes can be neglected completely (except in brunch choosing across singularities) in the computation of the wavepacket and mean-flow response. 

\begin{figure}
	\centering
	\includegraphics[width=1\textwidth]{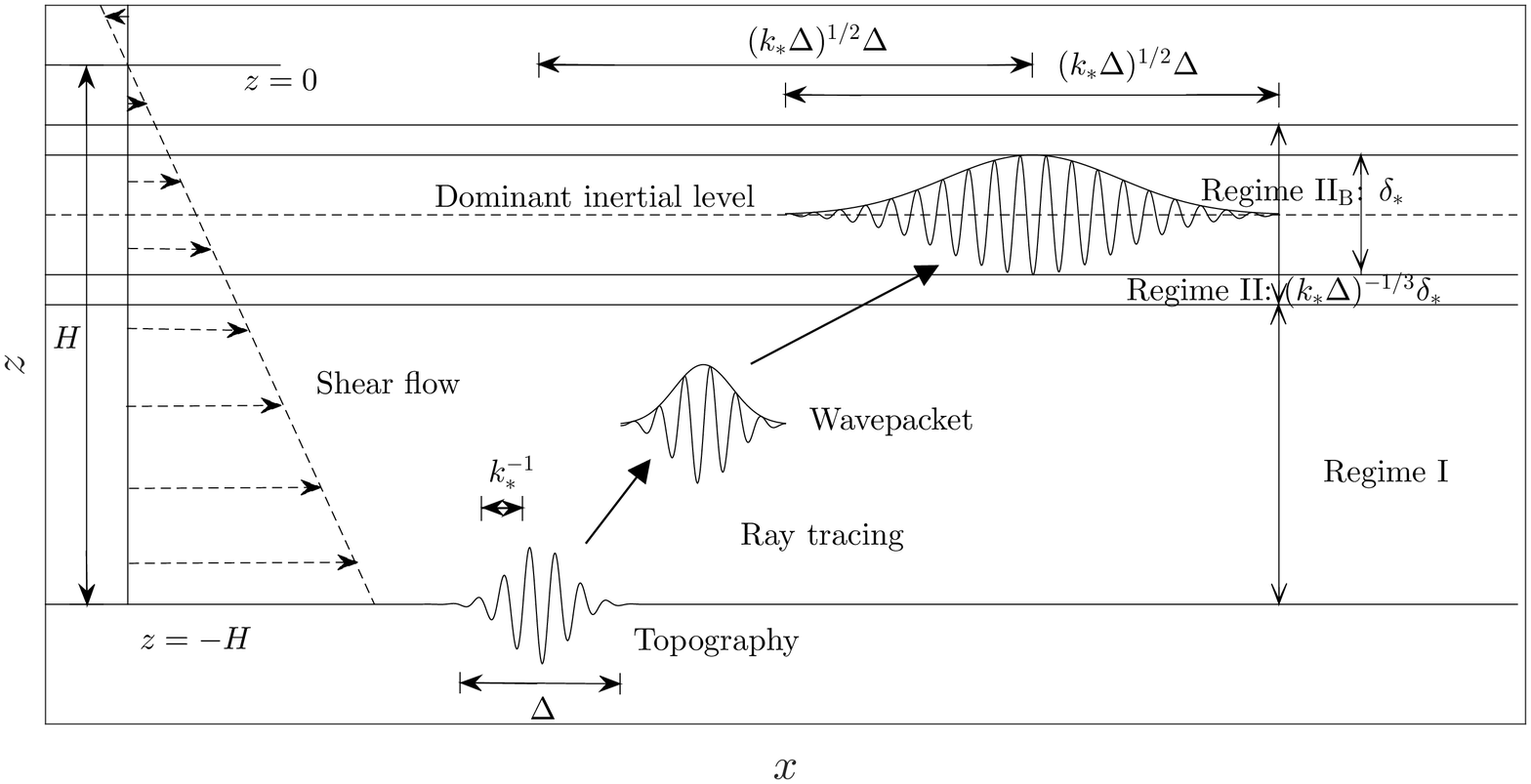}
	\caption{Scaling regimes for the mountain wavepacket: the  wavepacket is generated by a backsheared flow  over a two-scale topography, with a scale separation characterised by $k_*\Delta \ll 1$. Two distinguished regimes are found: Regime I and Regime II corresponding to distances from the dominant inertial level that are $O(H)$ and $O((k_*\Delta)^{-2/3}H)$, respectively. 
		The wave drag is localised in a region described by the limiting Regime II$_\mathrm{B}$, of {$O(\Delta_*)$} thickness around the dominant inertial level. This region is located  a large, $O((k_*\Delta)^{1/2}\Delta)$ distance downstream of the mountain and extending horizontally over an  $O((k_*\Delta)^{1/2}\Delta)$ scale.} \label{setup}
\end{figure}

By applying a steepest descent method, we obtain explicit approximations for the form of the wavepacket
in different regions characterised by their distance to the dominant inertial level, that is, the inertial level corresponding to the central wavenumber of the mountain profile. Our main conclusions concern the scaling of the wave solution and mean-flow forcing; they are indicated in Fig.\ \ref{setup}.
In Regime I, the wavepacket is sufficiently far away from the dominant inertial level that singular effects are not important. Standard ray-tracing results apply: the wavepacket resembles the topography, with envelope scale $\Delta$. 
In Regime II the wavepacket is close to the dominant inertial level and hence strongly affected by the inertial-level singularity in the vertical structure (\ref{vertial_str}) corresponding to a single wavenumber.
As a consequence, ray tracing does not apply, and the wavepacket has a characteristic streamwise scale $(k_*\Delta)^{1/3}\Delta$, much longer than the scale of the topography. 
In Regime II$_\mathrm{B}$, a subregime of Regime II, the wavepacket is closer still to the dominant inertial level and is absorbed. 
We pay special attention to this regime since it is relevant to the region where the horizontally-integrated EP flux varies vertically, leading to a drag on the mean flow. Qualitatively, our most important conclusion is about the location of this region, found to be an $O((k_*\Delta)^{1/2}\Delta)$ distance downstream of the mountain and to have an $O((k_*\Delta)^{1/2}\Delta)$ horizontal extent. Since $k_* \Delta \gg 1$, this makes it evident that mountain waves exert their drag far downstream of the mountain. This is in sharp contrast with their parameterisations in atmospheric models which are typically columnar, assuming that wave propagation is purely vertical and imposing their wave drag  right above the wave source  \citep[see][however]{Hash2008}.

%This study provides some indications on improving the topographic wave parameterization in numerical models.
%Parameterization of gravity waves as a subgrid-scale effect starts from \citet{Palm1986}, where momentum flux dissipates due to the gravity wave saturation and this columnar parameterization prevents information transferred between columns.
%Different dissipation mechanisms are applied in the literature, turbulence is the key dissipative effect for the parameterization of \citet{StLa2002}, and in \citet{Lott2013}'s columnar parameterization, same as in our setup, wave absorption is a result of singularity in the zero viscosity limit.
%The connection between neighbouring columns is accounted for in a ray-tracing based parameterization proposed by \citet{Hash2008}.
%Our results indicate that when the dissipation is a result of singularity in the zero viscosity limit the information exchange between columns should be taken into account -- IGWs force the mean flow of $O((k_*\Delta)^{1/2}\Delta)$ far away from the source and the force range is extended to a large range of $O((k_*\Delta)^{1/2}\Delta)$.  
%However, we also point out that ray tracing is not enough to accomplish this task due to its break down near inertial level singularity.

Using the form of the wave solution in Regime II$_\mathrm{B}$, we compute the far-field mean-flow response, taking advantage of the smallness of the Rossby number to use a quasi-geostrophic approximation. The vertical divergence of the EP flux, which controls the mean-flow response in this approximation, is localised in the thin inertial layer, with vertical scale $\delta_*$ defined in (\ref{ilscale}). The mean-flow response itself, however, has a large scale in both the horizontal and vertical directions because of the non-locality in the diagnostic relation between the mean potential vorticity and  mean streamfunction in the quasi-geostrophic approximation.
%By combining the EP flux conservation away from the inertial layer that the total EP flux is controlled at the bottom boundary topography, this non-local mean flow generation explains that the limit of $f \to 0$ does not bring about singular large-scale mean flow generation (Ref. \S \ref{SecEg}). 

An interesting feature of the mean-flow generation predicted by the Regime II$_\mathrm{B}$ approximation is that it is zero in the region with $X= k_*^{-1/2} \Delta^{-3/2} x < 0$ but jumps to an $X$-independent value for $X\geq 0$.
Implicit to this prediction is an assumption that $X=O(1)$, which gives the characteristic horizontal scale of the wavepacket in the inertial layer.  
The mean flow is in fact smoothly switched on over a shorter characteristic scale, specifically, $x/\Delta=O((k_*\Delta)^{1/3})$.
This result can be obtained using the Regime II approximations, but we do not carry out detailed calculations which are complicated by the presence of a branch cut in (\ref{ApprII}).

We emphasise that our results are more general than might seem at first glance. 
Our derivations are based on the distinguished scaling $\Ro = O((k_*\Delta)^{-1})$, $J = O(k_*\Delta)$, with $k_*\Delta\gg 1$, and $J h/H \ll 1$. In fact, their validity only requires that
\begin{equation}
J\frac{h}{H} \ll 1, \quad k_*\Delta \gg 1, \quad  J \gg 1, \quad \Ro \ll 1 \quad \textrm{and} \quad {r > 1} , \label{Para_Reg}
\end{equation}
corresponding to the validity of the hypotheses of (i) linear wave, (ii) scale separation of the mountain height, (iii) WKB scaling as in \cite{Lott2012}, (iv) quasi-geostrophic mean flow, and (v) the dominant inertial level located above  ground, ensuring that the argument of the square root in (\ref{EPfor_dim}) is positive. If the conditions in (\ref{Para_Reg}) are satisfied but the scaling differs from the assumed distinguished scaling, e.g.\ because $J \gg k_* \Delta$, our results continue to hold and could in fact be simplified by taking into account the existence of additional small parameters, such as $\Delta k_* / J$ in our example.

We conclude by assessing the validity of the assumption of infinitesimally small viscosity, which leads to a vertical scale of wave absorption given by the inertial-layer thickness (\ref{ilscale}).
%Here we provide an order-of magnitude comparison between the strength of inertial level singularity and viscosity. 
For a finite viscosity, the viscous vertical scale $\delta_\mathrm{d}=(\nu N^2/k_*)^{1/3}/\Lambda$ is found by considering the Taylor--Goldstein equation with dissipation, $D_t^2 \nabla^2 w + N^2 w_{xx} = \nu D_t \nabla^4 w$, with $D_t = \Lambda z \partial_x$.
The ratio $\delta_\mathrm{d}/\delta_*$ which measures the relative strength of viscosity and rotation in setting up the wave vertical scale is then found as $\delta_\mathrm{d}/\delta_*= \Delta k_*^{5/3} (\nu N^2)^{1/3} /f$, independent of the shear $\Lambda$ and relatively insensitive to the value of $\nu$.
Taking the atmospheric values $\nu=10^{-8}\,\mathrm{m^2s^{-1}}$, $N=10^{-2}\,\mathrm{s^{-1}}$, $f=10^{-4}\,\mathrm{s^{-1}}$, $k=10^{-3}\,\mathrm{m^{-1}}$ and $\Delta=10^4 \, \mathrm{m}$ as an illustration, we compute $\delta_\mathrm{d}/\delta_*=10^{-1}$, indicating a dominance of the rotation effects considered in this paper.

\vspace{1em}
\noindent{\bf Acknowledgements.} This research is funded by the UK Natural Environment Research Council (grant NE/J022012/1). J.-H.\ X.\ acknowledges financial support from the Centre for Numerical Algorithms and Intelligent Software (NAIS).

\appendix

\section{Details of wave solution} \label{appendix1}

In this Appendix we provide details of the derivation of the saddle-point approximations to the wavepacket in \S\ref{wave_solu}. 

\subsection{Regime I}\label{RegimeI}
In this section we consider altitudes well below the inertial level, corresponding to the scaling (\ref{z_exp}) for $z$  with $\alpha = 0$, that is, to
\begin{equation}
\zeta_* = -\frac{k_*\Lambda z}{f} = 1 + Z_{\mathrm{I}}, \quad \textrm{with} \ \ Z_\mathrm{I}=O(1).
\label{zzz}
\end{equation}
As discussed in \S\ref{sec:saddles}, the associated distinguished regime is obtained by considering the scaling (\ref{k_exp}) for the wavenumber $k$. 
Substituting (\ref{k_exp}) and (\ref{zzz}) into (\ref{w}), we obtain 
\begin{align} \label{wI}
w_1 &= \frac{\ii U_{\mathrm{b}}hk_*\Delta}{\epsilon\sqrt{2\pi}} \frac{k_*}{\epsilon}\br{\frac{\epsilon}{k_*\Delta}}^\beta
\ex^{-\br{ y - J \nu_* k_*^{-1} (1+\nu_*^2)^{-1/2} D_\mathrm{I} }^2/(2 \Delta^2)}
\\
&\times \int_{-\infty}^{\infty}
\frac{(1+Z_{\mathrm{I}})}{r}\br{\frac{r^2-1}{2Z_{\mathrm{I}}+Z_{\mathrm{I}}^2}}^{1/4} \ex^{-\epsilon^{2\beta-2}(k_*\Delta)^{2-2\beta} K_{\mathrm{I}}^2/2
	+ \ii \epsilon^{\beta-1}k_*(k_*\Delta)^{-\beta} K (x-X_{\mathrm{cI}})  } \, \dd K_{\mathrm{I}}, \nonumber
\end{align}
where 
\begin{equation}
X_{\mathrm{cI}} = \frac{J}{k_*}\left( 
\sqrt{1+\nu_*^2}
\br{ 
	\frac{1+Z_{\mathrm{I}}}{\sqrt{2Z_{\mathrm{I}}+Z_{\mathrm{I}}^2}} - \frac{r}{\sqrt{r^2 - 1}} }
-\frac{\nu_*^2}{\sqrt{1+\nu_*^2}}D_\mathrm{I}
\right)
\end{equation}
with $D_\mathrm{I}=\ln(1+Z_{\mathrm{I}}+(2Z_{\mathrm{I}}+Z_{\mathrm{I}}^2)^{1/2}) - \ln(r+(r^2-1)^{1/2})$. The expansion of $D(\zeta)$ results in the term $\ii \epsilon^{\beta-1}k_*(k_*\Delta)^{-\beta} K (x-X_{\mathrm{cI}})$ in the exponential, in agreement with  the scaling of $D(\zeta)$ in (\ref{d_exp2}).

A distinguished regime is obtained by balancing the arguments of the exponential in (\ref{wI}), corresponding to the choice $\beta=1$. With this, (\ref{wI}) can be integrated directly to find
\begin{equation}
w_1 \doteq \epsilon^{-1}h f (1+Z_{\mathrm{I}}) \br{\frac{r^2-1}{2Z_{\mathrm{I}}+Z_{\mathrm{I}}^2}}^{1/4}
\ex^{-\br{ y - J \nu_* k_*^{-1} (1+\nu_*^2)^{-1/2} D_\mathrm{I} }^2/(2 \Delta^2)}
\ex^{-(x-X_{\mathrm{cI}})^2/(2\Delta^2)},
\end{equation}
ignoring the phase factor for simplicity.

\subsection{Regime II}\label{RegimeII}

We now consider the wavepacket asymptotically close to the dominant inertial level, that is, for $\alpha > 0$ in the scaling (\ref{z_exp}) for $z$. In this case, the form of the expansion of $D(\zeta)$ in (\ref{d_exp}) depends on the smaller of $\alpha$ and $\beta$; therefore, a distinguished regime is naturally achieved with $\beta = \alpha$, leaving just the value of $\alpha$ to be determined.
With $\alpha=\beta$, $D(\zeta)$ expands as 
\begin{equation}
D(\zeta) \sim \br{\frac{\epsilon}{k_*\Delta}}^{\alpha/2} \sqrt{2(K_{\mathrm{II}}+Z_{\mathrm{II}})} -\ln(\zeta_\mathrm{b}+\sqrt{\zeta_\mathrm{b}^2-1}) - \br{\frac{\epsilon}{k_*\Delta}}^{\alpha} \frac{r}{\sqrt{r^2-1}} K_{\mathrm{II}}. \label{D2}
\end{equation}
Substituting this and the expansions (\ref{z_exp}) and (\ref{k_exp})  for $z$ and $k$  into (\ref{w}) leads to
\begin{align}
w_1 &= \frac{\ii U_{\mathrm{b}}hk_*^2\Delta}{2^{1/4}\epsilon^2\sqrt{2\pi}}\br{\frac{\epsilon}{k_*\Delta}}^{\alpha-1/4} \frac{(\zeta_\mathrm{b}^2-1)^{1/4}}{\zeta_\mathrm{b}}
\ex^{-\left( y - J \nu_* k_*^{-1} (1+\nu_*^2)^{-1/2} D_\mathrm{II} \right)^2/(2 \Delta^2)} \nonumber \\
&\times \int_{-Z_{\mathrm{II}}}^{\infty} \frac{1}{(K_{\mathrm{II}}+Z_{\mathrm{II}})^{1/4}} \ex^{-\epsilon^{2\alpha-2}(k_*\Delta)^{2-2\alpha} K_{\mathrm{I}}^2/2} 
\ex^{ -\ii \epsilon^{\alpha/2-1}(k_*\Delta)^{-\alpha/2}J(1+\nu_*^2)^{1/2} (2(K_{\mathrm{II}}+Z_{\mathrm{II}}))^{1/2} }  \label{wIII} \\
&\times \, \ex^{ \ii \epsilon^{\alpha-1} K_{\mathrm{II}} \br{ k_*(k_*\Delta)^{-\alpha} x+ 
		J(k_*\Delta)^{-\alpha}\left(
		(1+\nu_*^2)^{1/2} r \left(r^2-1\right)^{1/2} + 
		\nu_*^2 (1+\nu_*^2)^{-1/2}D_\mathrm{II}
		\right)
	} } \dd K_{\mathrm{II}}, \nonumber
	\end{align}
	where $D_\mathrm{II} = -\ln(r+\sqrt{r^2-1})$.
	A distinguished regime is obtained by balancing the arguments of the first two exponentials, leading to
	%\begin{equation}
	%\epsilon^{2\alpha-2}(k_*\Delta)^{2-2\alpha} K_{\mathrm{II}}^2/2 \sim
	%\ii \epsilon^{\alpha/2-1}(k_*\Delta)^{-\alpha/2}J\sqrt{1+\nu_*^2} \sqrt{2(K_{\mathrm{II}}+Z_{\mathrm{II}})},
	%\label{balance}
	%\end{equation}
	%from which we get
	$\alpha=2/3$. The third exponential contributes to the same order as the other two when  $x$ is suitably rescaled. 
	Substituting $\alpha = 2/3$ into (\ref{wIII}) we obtain
	\begin{align}
	w_1 &\doteq \frac{\ii U_{\mathrm{b}}hk_*^2\Delta}{2^{1/4}\sqrt{2\pi}\epsilon^{19/12} (k_*\Delta)^{5/12}}
	g(y)  \label{wII}  \\
	&\ttimes \int_{-\infty}^{\infty} \frac{1}{(K_{\mathrm{II}}+Z_{\mathrm{II}})^{1/4}} 
	\ex^{ \epsilon^{-2/3}(k_*\Delta)^{2/3}\br{ -{K_{\mathrm{II}}^2}/{2} - \ii J(k_*\Delta)^{-1/3}\sqrt{1+\nu_*^2}\sqrt{2(K_{\mathrm{II}}+Z_{\mathrm{II}})} + \ii K_{\mathrm{II}} X_{\mathrm{II}}}  } \dd K_{\mathrm{II}}, \nonumber
	\end{align}
	where $g(y)$ is the Gaussian given in (\ref{g(y)}). Here, 
	\begin{equation}
	X_{\mathrm{II}}=\epsilon^{1/3}\br{k_*(k_*\Delta)^{-4/3} x + J(k_*\Delta)^{-4/3}\left(
		\frac{r \sqrt{1+\nu_*^2}}{\sqrt{r^2-1}} + 
		\frac{\nu_*^2}{\sqrt{1+\nu_*^2}}D_\mathrm{II}
		\right) }
	\end{equation}
	is assumed to be $O(1)$, thus indicating that the horizontal scale of the wavepacket in Regime II is larger by a factor 
	$O(k_*\Delta)^{1/3}$ than its scale $ \Delta$ in  Regime I.

	We can now apply the saddle point method to approximate (\ref{wII}) as
	\begin{eqnarray}
	w_1 &=& \frac{\ii U_{\mathrm{b}}hk_*^2\Delta}{2^{1/4}\epsilon^{5/4} (k_*\Delta)^{3/4}} q(K_\mathrm{IIs}) \sqrt{\frac{1}{p''(K_\mathrm{IIs})}}
	g(y) \ex^{(k_*\Delta)^{2/3}P(K_\mathrm{IIs})}, 
	\end{eqnarray}
	where 
	\begin{eqnarray}
	q(K_{\mathrm{II}}) &=& \frac{1}{(K_{\mathrm{II}}+Z_{\mathrm{II}})^{1/4}},\nonumber \\
	p(K_{\mathrm{II}}) &=& -\frac{K_{\mathrm{II}}^2}{2} - \ii J(k_*\Delta)^{-1/3}\sqrt{1+\nu_*^2}\sqrt{2(K_{\mathrm{II}}+Z_{\mathrm{II}})} + \ii K_{\mathrm{II}} X_{\mathrm{II}},
	\end{eqnarray}
	and $K_\mathrm{IIs}$ is the saddle point such that $p'(K_\mathrm{IIs})=0$, with the prime denoting the derivative. %\commentJV{The branch of the square root, that is, the choice of the Riemann surface, should be determined before the saddle point method is applied. How do we determine the branch of the square root in the integral for $w_1$ above?} 
	The expression of $p$ indicates that there are three saddle points, but only one is accessible by a steepest descent path that connects $-\infty$ to $\infty$. 
	%\commentJV{Do you agree with the above sentence? Is the argument to choose the saddle that the contours cannot be deformed to get to the other two saddles? I do not like the argument based on finding absurdly large values.}
	%\commentJHX{Agree. I feel that we have returned to our first explanation on how to choose saddle point.}
	%\commentJV{Not sure what the following sentence mean: Numerically, we can find that except the chosen saddle point, the other two result in ridicule large or small values. How can anything be ridiculously large or small?}
	Note that the asymptotics of (\ref{ApprII}) for large $Z_{\mathrm{II}}$ matches the asymptotics of (\ref{wI}) for small $Z_{\mathrm{I}}$. This confirms that there is no distinguished regime between Regimes I and II. 
	The matching is also observed in Fig.\ \ref{FigWave}(c).
	
	\subsection{Regime II$_\textrm{B}$} \label{RegimeIIB}
	The previous two sections examined the two distinguished regimes of wave propagation. Here, we concentrate on the behaviour of the solution in the inertial layer of characteristic thickness $\delta_*$ as estimated in (\ref{ilscale}). 
	This defines Regime II$_\textrm{B}$, a subregime of Regime II characterised by  $\alpha=1$ and thus 
	\begin{equation}
	\zeta_* = 1 + \br{\frac{\epsilon}{k_*\Delta}}Z, \label{z_exp_IIB}
	\end{equation}
	with $Z=O(1)$. An argument analogous to that used for Regime II in \S \ref{RegimeII} then shows that the  wavenumber should be scaled as 
	\begin{equation}
	k=\epsilon^{-1}k_*\br{ 1+\frac{\epsilon}{k_*\Delta}K }, \label{k_exp_IIB}
	\end{equation}
	with $K=O(1)$.
	In principle we can deduce the form of $w_1$ in this regime  from the Regime-II result (\ref{ApprII}).
	However, it is simpler and more illuminating to work out this form directly from the definition of $w_1$.
	
	Introducing the scalings  (\ref{z_exp_IIB})--(\ref{k_exp_IIB}) into (\ref{w}), we obtain
	\begin{eqnarray}
	w_1 \!&=&\! \frac{\ii U_{\mathrm{b}}hk_*(k_*\Delta)^{1/4}}{2^{3/4}\pi^{1/2}\epsilon^{5/4}}
	g(y) \nonumber \\
	&&\times \int_{-\infty}^{\infty} \frac{\ex^{-\frac{K^2}{2}}}{(K+Z)^{1/4}}
	\ex^{ \epsilon^{-1/2}(k_*\Delta)^{1/2}\br{ - \ii J(k_*\Delta)^{-1}\sqrt{1+\nu_*^2}\sqrt{2(K+Z)} + \ii K X}  } \dd K,
	\label{w1IIB}
	\end{eqnarray}
	%With the scaling, the argument of the exponential in (\ref{w}) has the two terms
	%\begin{equation}
	%(k_*\Delta)^{2-2\alpha} K^2/2 \ll
	%\ii \epsilon^{-1/2}(k_*\Delta)^{-1/2}J\sqrt{1+\nu_*^2} \sqrt{2(K+Z)} \label{StaPha}
	%\end{equation}
	%(see (\ref{balance})). The second term dominates; 
	Since the argument of the dominant exponential is purely imaginary,  the stationary phase method can be applied in place of the more general saddle-point method. The stationary point is readily found as
	\begin{equation}
	K_\mathrm{s} = \frac{J^2\br{1+\nu_*^2}}{2(k_*\Delta)^{2}X^2}-Z. \label{SP}
	\end{equation}
	Using this, the stationary phase approximation of (\ref{w1IIB}) is found as
	\begin{eqnarray}
	w_1 &\dot=& \frac{\ii h f J^{1/2}\br{1+\nu_*^2}^{1/4} \br{r^2 -1}^{1/4}}{(k_*\Delta)^{1/2}}
	\ex^{-\br{ y - J \nu_* k_*^{-1} (1+\nu_*^2)^{-1/2} D_\mathrm{II} }^2/(2 \Delta^2)} \nonumber \\
	%\ex^{ -\frac{1}{2}\br{\frac{J^2\br{1+\nu_*^2}}{2(k_*\Delta)^{2}X^2}-Z}^2 }. 
	&\times& \ex^{ -\left(a^2/X^2-Z\right)^2/2} \, \ex^{-\ii (k_* \Delta)^{1/2} \left(J^2 (1+\nu_*^2)/(2 (k_* \Delta)^2 X) + Z X \right)}  
	\label{fullw1IIB}
	\end{eqnarray}
	where 
	\begin{equation}
	X=\br{k_*(k_*\Delta)^{-3/2} x - J(k_*\Delta)^{-3/2}\left(
		\frac{r \sqrt{1+\nu_*^2}}{\sqrt{r^2-1}} + 
		\frac{\nu_*^2}{\sqrt{1+\nu_*^2}}D_\mathrm{II}
		\right) },
	\label{X}
	\end{equation}
	$a$ is defined in (\ref{a}), and we have ignored the rapidly varying phase and have set $\epsilon=1$. 

	\section{EP flux at the ground} \label{app:bottom}
	
	To obtain the integrated EP flux at the bottom of the domain, we use that
	\begin{equation}
	\overline{F}_{\mathrm{tot}}  = 4\pi^2 \int \int F(\bk) \, \dd k \dd l,
	\end{equation}
	where $F(\bk)$ on the right-hand side is the plane-wave expression for the EP flux in (\ref{EPk}) and the factor $4\pi^2$ arises in the transformation of the surface integral into a Fourier integral. The form of $F(\bk)$ is dominated by its first term; further, it follows from (\ref{vertial_str0}) that
	\begin{equation}
	\hat{w}_\zeta \sim  -\ii \frac{J\sqrt{1+\nu^2}}{{\sqrt{\zeta^2-1}}}\hat{w}.
	\end{equation}
	The integrated EP flux is therefore approximated as 
	\begin{equation}
	\begin{aligned}
	\overline{F}_{\mathrm{tot}} 
	& =  \frac{4\pi^2 J \Lambda \sqrt{r^2-1}}{f r^2\sqrt{1+\nu_*^2}} \iint |w(\bk)|^2 \, \dd k \dd l =  \frac{4\pi^2 J \Lambda \sqrt{r^2-1}}{f r^2\sqrt{1+\nu_*^2}} \frac{k_*^2 U_{\mathrm{b}}^2 h^2 \Delta^4}{4\pi^2}\frac{\pi}{\Delta^2} \\
	& = \frac{\pi J \Delta^2 \Lambda h^2f \br{r^2-1}^{1/2} }{\br{1+\nu_*^2}^{1/2}},
	\end{aligned}
	\end{equation} 
	in agreement with (\ref{EPt1}).
	
{
\section{Details of the mean-flow response} \label{SecEg}

We give the explicit solution of the linear system (\ref{PVbc}), (\ref{conti_cond}) and (\ref{matching}) for $\hat C_3$, $\hat C_4$ and $\hat C_5$. Defining
\[
L = \frac{N}{f} (H+z_*)= J(r - 1)/k_* \quad \textrm{and} \quad 
D = \frac{\sqrt{\pi}J^2 (k_* \Delta) \Lambda^2 h^2  (r^2 -1)^{1/2}}{f (1+\nu_*^2)^{1/2}},
\]
we rewrite  (\ref{PVbc}), (\ref{conti_cond}) and (\ref{matching}) as
\begin{subequations}
	\begin{align}
	-\hat C_4\ex^{L |l|} + \hat C_5 \ex^{-L |l|} &= 0, \\
	\hat C_4+ \hat C_5- \hat C_3 &= 0, \\
	\hat C_4- \hat C_5- \hat C_3  &= D\frac{\widehat{\partial_y g^2}}{|l|}.
	\end{align}
\end{subequations}
Solving gives
\begin{equation}
\hat C_3 = -\frac{D}{2} \left( 1 + \ex^{-2L|l|} \right) \frac{\widehat{\partial_y g^2}}{|l|}, \quad
\hat C_4 = -\frac{D}{2} \ex^{-2L|l|} \frac{\widehat{\partial_y g^2}}{|l|} \quad \textrm{and}
\quad 
\hat C_5 = -\frac{D}{2}  \frac{\widehat{\partial_y g^2}}{|l|}.
\end{equation}
The flow displayed in Figure \ref{mean} is obtained for the parameter values $J=k_*\Delta$, $l_*=0$ and $r=2$. 
In this case, $L = \Delta$, $D=\sqrt{3 \pi} J^3 \Lambda^2 h^2 / f$ and $\widehat {\partial_y g^2} = - \ii \Delta l \exp(-\Delta^2 l^2/4)/\sqrt{2}$ (see (\ref{g(y)})). 
}

%
%\bibliographystyle{jfm}
%\bibliography{Topo_wave}

\begin{thebibliography}{25}
	\expandafter\ifx\csname natexlab\endcsname\relax\def\natexlab#1{#1}\fi
	\def\au#1{#1} \def\ed#1{#1} \def\yr#1{#1}\def\at#1{#1}\def\jt#1{\textit{#1}}
	\def\bt#1{#1}\def\bvol#1{\textbf{#1}} \def\vol#1{#1} \def\pg#1{#1}
	\def\publ#1{#1}\def\arxiv#1{#1}\def\org#1{#1}\def\st#1{\textit{#1}}
	
	\bibitem[Alexander {\em et~al.\/}(2010)Alexander, Geller, McLandress,
	Polavarapu, Preusse, Sassi, Sato, Eckermann, Ern, Hertzog, Kawatani, Pulido,
	Shaw, Sigmond, Vincent \& Watanabe]{Alex2010}
	{\sc \au{Alexander, M.~J.}, \au{Geller, M.}, \au{McLandress, C.},
		\au{Polavarapu, S.}, \au{Preusse, P.}, \au{Sassi, F.}, \au{Sato, K.},
		\au{Eckermann, S.}, \au{Ern, M.}, \au{Hertzog, A.}, \au{Kawatani, Y.},
		\au{Pulido, M.}, \au{Shaw, T.~A.}, \au{Sigmond, M.}, \au{Vincent, R.} \&
		\au{Watanabe, S.}} \yr{2010}  \at{Recent developments in gravity-wave effects
		in climate models and the global distribution of gravity-wave momentum flux
		from observations and models}.  \jt{Quart. J. R. Met. Soc.}
	\bvol{136}~(650),  \pg{1103--1124}.
	
	\bibitem[Andrews \& McIntyre(1976)]{Andr1976}
	{\sc \au{Andrews, D.~G.} \& \au{McIntyre, M.~E.}} \yr{1976}  \at{Planetary
		waves in horizontal and vertical shear: the generalized {E}liassen-{P}alm
		relation and the mean zonal acceleration}.  \jt{J. Atmos. Sci.}  \bvol{33},
	\pg{2031--2048}.
	
	\bibitem[Andrews \& McIntyre(1978)]{Andr1978}
	{\sc \au{Andrews, D.~G.} \& \au{McIntyre, M.~E.}} \yr{1978}  \at{Generalized
		{E}liassen-{P}alm and {C}harney-{D}razin theorems for waves on axisymmetric
		mean flows in compressible atmospheres}.  \jt{J. Atmos. Sci.}  \bvol{35},
	\pg{175--185}.
	
	\bibitem[Booker \& Bretherton(1967)]{Book1967}
	{\sc \au{Booker, J.~R.} \& \au{Bretherton, F.~P.}} \yr{1967}  \at{The critical
		layer for internal gravity waves in a shear flow}.  \jt{J. Fluid Mech.}
	\bvol{27},  \pg{513--539}.
	
	\bibitem[Boyd(1976)]{Boyd1976}
	{\sc \au{Boyd, J.~P.}} \yr{1976}  \at{The noninteraction of waves with the
		zonally averaged flow on a spherical earth and the interrelationships on eddy
		fluxes of energy, heat and momentum}.  \jt{J. Atmos. Sci.}  \bvol{33},
	\pg{2285--2291}.
	
	\bibitem[Bretherton(1969{\natexlab{{\em a\/}}})]{Bret1969}
	{\sc \au{Bretherton, F.}} \yr{1969{\natexlab{{\em a\/}}}}  \at{On the mean
		motion induced by internal gravity waves}.  \jt{J. Fluid Mech.}  \bvol{36},
	\pg{758--803}.
	
	\bibitem[Bretherton(1966)]{Bret1966}
	{\sc \au{Bretherton, F.~P.}} \yr{1966}  \at{The propagation of groups of
		internal gravity waves in a shear flow}.  \jt{Quart. J. Roy. Meteor. Soc.}
	\bvol{92},  \pg{466--480}.
	
	\bibitem[Bretherton(1969{\natexlab{{\em b\/}}})]{Bret1969b}
	{\sc \au{Bretherton, F.~P.}} \yr{1969{\natexlab{{\em b\/}}}}  \at{Momentum
		transport by gravity waves}.  \jt{Quart. J. Roy. Meteor. Soc.}  \bvol{95},
	\pg{213--243}.
	
	\bibitem[Charney \& Drazin(1961)]{Char1961}
	{\sc \au{Charney, J.~G.} \& \au{Drazin, P.~G.}} \yr{1961}  \at{Propagation of
		planetary-scale disturbances from the lower into the upper atmosphere}.
	\jt{J. Geophys. Res.}  \bvol{66},  \pg{83--109}.
	
	\bibitem[{Edmon} {\em et~al.\/}(1980){Edmon}, {Hoskins} \&
	{McIntyre}]{Edmo1980}
	{\sc \au{{Edmon}, Jr., H.~J.}, \au{{Hoskins}, B.~J.} \& \au{{McIntyre}, M.~E.}}
	\yr{1980}  \at{Eliassen-{P}alm cross sections for the troposphere.}  \jt{J.
		Atmos. Sci.}  \bvol{37},  \pg{2600--2616}.
	
	\bibitem[Eliassen \& Palm(1961)]{Elia1961}
	{\sc \au{Eliassen, A.} \& \au{Palm, E.}} \yr{1961}  \at{On the transfer of
		energy in stationary mountain waves}.  \jt{Geopy. Publ.}  \bvol{22},
	\pg{1--23}.
	
	\bibitem[Fritts \& Alexander(2003)]{Frit2003}
	{\sc \au{Fritts, D.~C.} \& \au{Alexander, M.~J.}} \yr{2003}  \at{Gravity wave
		dynamics and effects in the middle atmosphere}.  \jt{Rev. Geophys.}
	\bvol{41},  \pg{1003}.
	
	\bibitem[Hasha {\em et~al.\/}(2008)Hasha, B\"uhler \& Scinocca]{Hash2008}
	{\sc \au{Hasha, A.}, \au{B\"uhler, O.} \& \au{Scinocca, J.}} \yr{2008}
	\at{Gravity wave refraction by three-dimensionally varying winds and the
		global transport of angular momentum}.  \jt{J. Atmos. Sci.}  \bvol{65},
	\pg{2892--2906}.
	
	\bibitem[Jones(1967)]{Jones1967}
	{\sc \au{Jones, W.~L.}} \yr{1967}  \at{Propagation of internal gravity waves in
		fluids with shear flow and rotation}.  \jt{J. Fluid Mech.}  \bvol{30},
	\pg{439--448}.
	
	\bibitem[Lott {\em et~al.\/}(2015)Lott, Millet \& Vanneste]{Lott2015}
	{\sc \au{Lott, F.}, \au{Millet, C.} \& \au{Vanneste, J.}} \yr{2015}
	\at{Inertia-gravity waves in inertially stable and unstable shear flows}.
	\jt{J. Fluid Mech.}  \bvol{775},  \pg{223--240}.
	
	\bibitem[Lott {\em et~al.\/}(2010)Lott, Plougonven \& Vanneste]{Lott2010}
	{\sc \au{Lott, F.}, \au{Plougonven, R.} \& \au{Vanneste, J.}} \yr{2010}
	\at{Gravity waves generated by sheared potential vorticity anomalies}.
	\jt{J. Atmos. Sci.}  \bvol{67},  \pg{157--170}.
	
	\bibitem[Lott {\em et~al.\/}(2012)Lott, Plougonven \& Vanneste]{Lott2012}
	{\sc \au{Lott, F.}, \au{Plougonven, R.} \& \au{Vanneste, J.}} \yr{2012}
	\at{Gravity waves generated by sheared three-dimensional potential vorticity
		anomalies}.  \jt{J. Atmos. Sci.}  \bvol{69},  \pg{2134--2151}.
	
	\bibitem[Martin \& Lott(2007)]{Mart2007}
	{\sc \au{Martin, A.} \& \au{Lott, F.}} \yr{2007}  \at{Synoptic responses to
		mountain gravity waves encountering directional critical levels}.  \jt{J.
		Atmos. Sci.}  \bvol{64},  \pg{828--848}.
	
	\bibitem[Nikurashin \& Ferrari(2011)]{Niku2011}
	{\sc \au{Nikurashin, M.} \& \au{Ferrari, R.}} \yr{2011}  \at{Global energy
		conversion rate from geostrophic flows into internal lee waves in the deep
		ocean}.  \jt{Geophys. Res. Lett.}  \bvol{38},  \pg{L08610}.
	
	\bibitem[Nikurashin \& Ferrari(2013)]{Niku2013}
	{\sc \au{Nikurashin, M.} \& \au{Ferrari, R.}} \yr{2013}  \at{Overturning
		circulation driven by breaking internal waves in the deep ocean}.
	\jt{Geophys. Res. Lett.}  \bvol{41},  \pg{3133--3137}.
	
	\bibitem[Scott {\em et~al.\/}(2011)Scott, Goff, Garabato \& Nurser]{Scot2011}
	{\sc \au{Scott, R.~B.}, \au{Goff, J.~A.}, \au{Garabato, A. C.~Naveira} \&
		\au{Nurser, A. J.~G.}} \yr{2011}  \at{Global rate and spectral
		characteristics of internal gravity wave generation by geostrophic flow over
		topography}.  \jt{J. Geophys. Res.}  \bvol{116}.
	
	\bibitem[Shutts(1995)]{Shut1995}
	{\sc \au{Shutts, G.}} \yr{1995}  \at{Gravity-wave drag parametrization over
		complex terrain: The effect of critical-level absorption in directional
		wind-shear}.  \jt{Quart. J. Roy. Meteor. Soc.}  \bvol{121},  \pg{1005--1021}.
	
	\bibitem[Shutts(2001)]{Shut2001}
	{\sc \au{Shutts, G.}} \yr{2001}  \at{A linear model of back-sheared flow over
		an isolated hill in the presence of rotation}.  \jt{J. Atmos. Sci.}
	\bvol{58},  \pg{3293--3311}.
	
	\bibitem[Shutts(2003)]{Shut2003}
	{\sc \au{Shutts, G.}} \yr{2003}  \at{Inertia gravity wave and neutral {E}ady
		wave trains forced by directionally sheared flow over isolated hills}.
	\jt{J. Atmos. Sci.}  \bvol{60},  \pg{593--606}.
	
	\bibitem[Yamanaka \& Tanaka(1984)]{Yama1984}
	{\sc \au{Yamanaka, M.~D.} \& \au{Tanaka, H.}} \yr{1984}  \at{Propagation and
		breakdown of internal inertio-gravity waves near critical levels in the
		middle atmosphere.}  \jt{J. Meteor. Soc. Japan}  \bvol{62},  \pg{1--16}.
	
\end{thebibliography}
\end{document}